\documentclass[preprintnumbers,nofootinbib,showkeys,showpacs,amsmath,amssymb]{revtex4}
\usepackage{amsmath,amssymb,graphics,epsfig,subfigure}
\usepackage{color}
\usepackage{multirow}
\usepackage{booktabs}
\usepackage{changes}
\usepackage{footnote}

\usepackage{hyperref}
\hypersetup{colorlinks=true,linkcolor=blue,citecolor=magenta}

\begin{document}
	\renewcommand{\baselinestretch}{1.15}
	
\title{Thermodynamical Topology of Quantum BTZ Black Hole}
	
\preprint{}
	
\author{Shan-Ping Wu, Shao-Wen Wei \footnote{Corresponding author. E-mail: weishw@lzu.edu.cn}}

\affiliation{$^{1}$Lanzhou Center for Theoretical Physics, Key Laboratory of Theoretical Physics of Gansu Province, and Key Laboratory of Quantum Theory and Applications of MoE, Lanzhou University, Lanzhou, Gansu 730000, China,\\
		$^{2}$Institute of Theoretical Physics $\&$ Research Center of Gravitation, Lanzhou University, Lanzhou 730000, People's Republic of China}

\begin{abstract}
Among the study of black hole thermodynamics, topology offers a novel approach and perspective for classifying black hole systems. In this work, we explore the thermodynamical topology of the quantum BTZ black hole by employing the concept of the generalized free energy. To fully characterize the thermodynamics, we introduce two distinct topological numbers. The first one is determined by an expression, denoted by $z$, derived from the free energy. Although it can provide us with some local physical explanations, sufficient physical significance still lacks from a global perspective. On the other hand, the second topological number is based on the entropy expression of the generalized free energy, leading to a more meaningful interpretation of its physical implications. This result highlights the natural choice of entropy as the domain variable for the generalized free energy. Regarding the second topological number, our analysis reveals a topological transition that is associated with the thermodynamical stability of the ``cold" black hole state of the quantum BTZ black hole. And the thermodynamical topology of BTZ black hole and quantum BTZ black hole can be different, which implies a significant impact of quantum effects on thermodynamics. Furthermore, our study suggests the existence of topological numbers beyond the conventional values of $\pm 1,0$.
\end{abstract}
	
\keywords{Quantum BTZ black hole, thermodynamics, topology}
	\pacs{04.70.-s, 04.70.Dy, 02.40.Re}
	
	\maketitle

\section{Introduction}\label{intro}

The black hole, as a distinctive spacetime structure, continues to be a focus of numerous advanced studies in the field of gravity. Significantly, recent observations showed the evidence that the black holes exist in our universe~\cite{LIGOScientific:ObservationGravitationalWaves,EventHorizonTelescope:2019ImagingSuperBH,EventHorizonTelescope:2019ShadowBH,LIGOScientific:2020BinaryBHMerger}. When quantum theory is taken into account, one intriguing aspect of black holes, the Hawking radiation~\cite{Hawking:1975ParticleCreation}, emerges. Further treating the surface gravity and area of the black hole horizon as the temperature and entropy, black hole systems are regarded as thermodynamical systems~\cite{Hawking:1976BHTherm,Bardeen:1976FourLaws,Gibbons:1976ActionIntegral}. Meanwhile, the entropy-area law implies the holographic nature of the quantum gravity~\cite{tHooft:1993DimensionReduction,Susskind:1994WorldHologram,Maldacena:1997LargeN,Ryu:2006EntanglementEntropy}. Consequently, the thermodynamics of black holes serves as a manifestation of certain quantum gravitational effects.

Similar to the general thermodynamical systems, it is conceivable that black holes possess phases and can undergo phase transitions. In particular, in the AdS space, there is the Hawking-Page phase transition between the pure radiation phase and the stable large Schwarzschild black hole phase~\cite{Hawking:1982ThermAdSBH}. Such phase transition can be used to interpret the transition between the confinement and unconfinement in gauge theory~\cite{Witten:1998AdSThermalConfine}. For the charged or rotating AdS black holes, a first order phase transition between the small and large black hole phases was observed, and which is analogy to the liquid-gas system~\cite{Kubiznak:2012ChargedAdSBH,Li:2020RNAdS,Altamirano:2013KerrAdS,Yang:2021KerrAdSBH}. Moreover, black hole phase transition can also provide insights into their underlying microscopic configuration~\cite{Wei:2015InsightMicroStruct}.

To better understand and characterize distinct black hole systems, the concept of thermodynamical topology has been well introduced~\cite{Wei:2022BHasTopDefect}. This approach relies on the asymptotic behavior of the generalized free energy of black holes, allowing for the classification of almost all black hole systems into three distinct categories associated with topological numbers: $+1$, $0$, and $-1$. These numbers also correspond to the difference between the numbers of stable and unstable phases exhibited by the black holes. For examples, the Schwarzschild black holes, charged black holes, and charged AdS black holes are assigned topological number of $-1$, 0, 1, respectively. Furthermore, the thermodynamical topology of various other black holes have been studied~\cite{Wu:2022TopoRotate,Wu:2023TopoAdSRotat,Du:2023BTZ}. The utilization of topological numbers to classify the thermodynamics of black holes showcases their remarkable capability, offering potential insights into the nature of black hole systems. This approach holds promise for advancing our understanding of black holes. Moreover, topology can also serve as a valuable tool for describing the critical points of phase transitions in black hole systems~\cite{Bai:2022TopoLovelock,Fan:2022TopoInterpretation,Chen:2023BornInfeldBHflatSpace,Yerra2024jh,Yerra2022}. Additionally, there are other ways to thermodynamical topology~\cite{Wei:2021TopologyBH,Yerra:2022HawkingPage,Fang:2022RevisitingThermTopo}.

In recent developments, a three-dimensional quantized black hole, known as the quantum BTZ (quBTZ) black hole, has been proposed~\cite{Emparan:1999brane2,Emparan:1999ExactBHBrane,Emparan:2002qBHAdSbraneWorld,Emparan:2020quBTZ}. This intriguing black hole arises from the introduction of an $AdS_3$ brane into an asymptotically $AdS_4$ spacetime, as described by the C-metric~\cite{Karch:2000Locallygravity}. The intersection of this brane with the bulk event horizon leads to the formation of a black hole on the $AdS_3$ brane. Within the framework of braneworld holography, the classical solution in the bulk is thought to relate with the quantum gravity and conformal field on the brane. Consequently, this black hole solution on the brane can be naturally regarded as inherently quantized~\cite{Emparan:2002qBHAdSbraneWorld}. Thus this intriguing connection between the classical bulk solution and quantum gravity on the brane sheds new light into the quantized black holes.

Similar to other black holes, the quBTZ black hole is also considered as a thermodynamical system, inheriting certain thermodynamical properties from its bulk counterparts. However, unlike the Hawking-Page phase transition observed in the BTZ black hole~\cite{Eune:2013HPinBTZBH}, the quBTZ black holes demonstrate a more interesting phase transition, the reentrant phase transition~\cite{Frassino:2023RPTQBTZ}. As the temperature increases from zero, the system undergoes a transition from a thermal AdS phase to a black hole phase, and eventually returns back to the thermal AdS phase. This fascinating behavior is a distinctive feature of the quBTZ black holes compared to their non-quantized counterparts. Furthermore, the origin and specific heats of quBTZ black holes have also been studied in the context of the extended thermodynamics~\cite{Antonia,Johnson:2023SHQBTZ}. The criticality and thermodynamic geometry are also considered~\cite{Mansoori}. These results shed light on the thermodynamical properties and behavior of the quBTZ black hole system.

Motivated by the referred distinctive properties, we, in this work, aim to study the thermodynamical topology for the quBTZ black holes. Our study reveals several remarkable characteristics of the quBTZ black hole thermodynamics that distinguishes from the classical BTZ black hole. Unlike its classical counterpart, the entropy of the quBTZ black holes is finite, and does not exhibit a monotonic behavior with respect to the black hole parameter $z$. This intriguing observation has prompted us to re-examine the thermodynamical topology, leading us to identify a topological transition associated with the quBTZ black holes. This differs from classical non-rotating BTZ black holes with fixed topological number $1$~\cite{Du:2023BTZ}. Furthermore, through the study of quBTZ black holes, we have gained a deeper understanding of the significance of topology in the context of black hole thermodynamics. This investigation allows us to explore the relationship between the thermodynamical properties of black holes and their underlying topological features, shedding light on the implications of topology in this fascinating field.

The paper is organized as follows: In Sec.~\ref{Sec_qBZTBlackHole}, we provide a brief introduction to the quantum BTZ black holes. Sec.~\ref{Sec_ThermTopology} is dedicated to the study of the quBTZ black hole's topology. We uncover that the critical boundary obtained from $\partial_z S = 0$ holds crucial importance for its physical interpretation. In Sec.~\ref{Sec_Revisit}, we reconstruct the vector mapping and obtain a new topology. To further illustrate our findings, in Sec.~\ref{Sec_Example}, we present comparative examples that analyze the thermodynamical topologies. Finally, we summarize and discuss the results in Sec.~\ref{Sec_Disscusion}.

\section{Quantum BTZ black holes}\label{Sec_qBZTBlackHole}

In the static asymptotically $AdS_4$ spacetime described by the C-metric, it is possible to introduce a $AdS_3$ brane. On this brane, the quBTZ black holes can be derived, and the corresponding metric reads~\cite{Emparan:2020quBTZ},
\begin{eqnarray}
	ds^2=-f(r)dt^2+\frac{dr^2}{f(r)}+r^2d\phi ^2,
\end{eqnarray}
with
\begin{eqnarray}
	f(r)=\frac{r^2}{\ell _{3}^{2}}-8\mathcal{G} _3M-\frac{\ell \mathcal{F} (M)}{r},
\end{eqnarray}
wherein $\mathcal{G}_3 = G_3/\sqrt{1+\ell^2/\ell_3^2}$ is renormalized gravitational constant, and $\ell$ is directly linked to the brane tension and the strength of back-reaction in the dual theory. As $\ell \rightarrow 0$, the brane approaches the $AdS_4$ boundary, and the brane tension diverges. In terms of the perspective on this brane, the limit $\ell \rightarrow 0$ also signifies the vanishing of the back-reaction from the conformal field theory, and quBTZ black hole will become classical BTZ black hole.

\begin{figure}
	\begin{center}
		{\includegraphics[width=7cm]{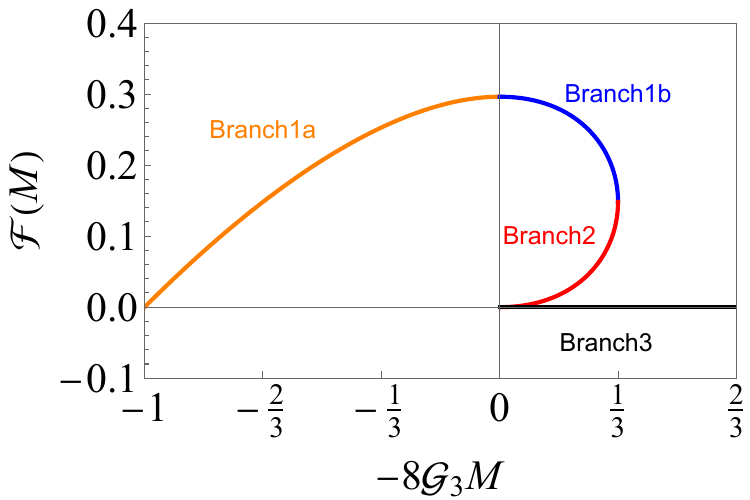}}
	\end{center}
	\caption{The holographic stress-energy function $\mathcal{F}(M)$. There are three branches of black hole. Branches 1 (1a, 1b) and 2 represent the quBTZ black hole solutions, and branch 3 represents the classical BTZ black hole.} \label{fig_F(M)}
\end{figure}

In the presence of quantum corrections, the holographic stress-energy function $\mathcal{F}(M)$ is complicated, but we can express it in term of the parameter $-\kappa x_1^2$ ($-1<-\kappa x_1^2<\infty$)
\begin{equation}
	M=\frac{1}{2\mathcal{G} _3}\frac{-\kappa x_{1}^{2}}{\left( 3-\kappa x_{1}^{2} \right) ^2},\quad \mathcal{F} \left( M \right) =8\frac{1-\kappa x_{1}^{2}}{\left( 3-\kappa x_{1}^{2} \right) ^3}.
	\label{eq_kx2_M_F(M)}
\end{equation}
A more explicit relationship is depicted in Fig.~\ref{fig_F(M)}, which showcases three distinct branches. Branch 1 (composed of branch 1a and branch 1b) and branch 2 correspond to the quBTZ black holes, as described by Eq.~\eqref{eq_kx2_M_F(M)}. However, branch 3 is for the black string in bulk and represents the classical BTZ black hole on the brane. For our purpose, we concentrate solely on branches 1 and 2, which characterize the quBTZ black holes.

In order to examine the black hole thermodynamics, two real and non-negative parameters~\cite{Emparan:2020quBTZ}
\begin{equation}
	z = \frac{\ell_{3}}{r_+ x_1},\quad	\nu =  \frac{\ell}{\ell_3},
\end{equation}
are introduced, where $r_+$ denotes the radius of the quBTZ black hole horizon. The following relation
\begin{equation}
	-\kappa x_{1}^{2}=\frac{1-\nu z^3}{z^2\left( 1+\nu z \right)},
\end{equation}
shows a one-to-one and smooth mapping between $0<z<\infty$ and $-1<-\kappa x_{1}^{2}<\infty$ for $\nu > 0$. Moreover, the thermodynamical quantities of the black holes can be expressed in terms of $z$ and $\nu$
\begin{eqnarray}
	M&=&\frac{1}{2\mathcal{G} _3}\frac{z^2(1-\nu z^3)(1+\nu z)}{(1+3z^2+2\nu z^3)^2}, \label{eq_energy}\\
	T&=&\frac{1}{2\pi \ell _3}\frac{z(2+3\nu z+\nu z^3)}{1+3z^2+2\nu z^3},\label{eq_temperature}\\
	S&=&\frac{\pi \ell _3}{\mathcal{G} _3}\frac{z}{1+3z^2+2\nu z^3}. \label{eq_entropy}
\end{eqnarray}
By differentiating $M$ and $S$ with respect to $z$, one can easily obtain the following first law of black hole thermodynamics
\begin{equation}
	dM  = T dS,\label{eq_FirstLawofTherm}
\end{equation}
for fixed $\ell_3$, $\nu$, and $\mathcal{G}_3$. This indicates that, similar to these black holes without quantum correction, the quBTZ black hole can also be treated as a thermodynamical system.

In order to well describe the thermodynamics, we show three characteristic values of $z$
\begin{eqnarray}
	z_1=\frac{1}{\nu ^{1/3}},\quad z_2=-\frac{1}{\nu}+2\sqrt{1+\frac{1}{\nu ^2}}\sin \left( \frac{1}{3}\alpha +\frac{\pi}{6} \right), 	\label{eq_z1z2}
	\\
	\hat{z}=\frac{1}{4\nu}\left( -1+\left( 8\nu ^2-1+4\nu \sqrt{4\nu ^2-1} \right) ^{-1/3}+\left( 8\nu ^2-1+4\nu \sqrt{4\nu ^2-1} \right) ^{1/3} \right),\label{eq_hatz}
\end{eqnarray}
where $\alpha = \arccos\left(1/\sqrt{1+\nu^2}\right)$. It is noteworthy that $z_1$ and $z_2$ are the solutions of $\partial_z {T} = 0$, and $\hat{z}$ is the solution of $\partial_{z}S = 0$ or $\partial_{z} M = 0$.

Through a straightforward analysis, it is easy to conclude that both $z_1$ and $z_2$ are greater than $\hat{z}$. Additionally, $\nu \geqslant 1$ ($\nu<1$) implies $z_2 \geqslant z_1$ ($z_2 < z_1$). As discussed in Ref.~\cite{Frassino:2023RPTQBTZ}, the ranges of $z$ for the ``cold" black hole, ``intermediate" black hole, and ``hot" black hole are listed in TABLE.~\ref{Table_ThermodynamicState}.

\begin{table}
	\centering
	\resizebox{12cm}{!}{
		\begin{tabular}{cccc}
			\hline
			\ \ \ \ \ \ Range \ \ \ \ \ \ & \ \ \ \ \ \ ``cold'' black hole \ \ \ \ \ \ &  \ \ \ \ \ \ ``intermediate'' black hole \ \ \ \ \ \ & \ \ \ \ \ \ ``hot'' black hole \ \ \ \ \ \  \\
			\hline
			$ \nu > 1 $  & $0<z<z_1 $ & $z_1<z<z_2$ & $z_2<z<\infty$ \\
			$ \nu = 1 $  & $0<z<z_1 $ & vanish ($z_1 = z_2$) & $z_1 < z < \infty$ \\
			$ \nu < 1 $  & $0<z<z_2 $ & $z_2<z<z_1$ & $z_1 < z < \infty$ \\
			\hline
	\end{tabular}}
	\caption{Thermodynamical states of quBTZ black holes with different parameter $\nu$.}\label{Table_ThermodynamicState}
\end{table}

\begin{figure}
	\begin{center}
		\subfigure[ \ $z-\mathcal{G}_3 S/(\pi \ell_3)$
		\label{FIGzS}]{\includegraphics[width=4.3cm]{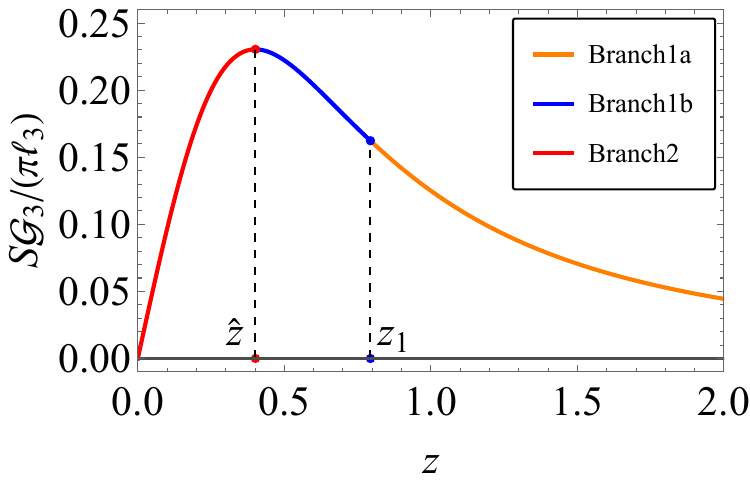}
			\label{FIG_ThermzS}}
		\subfigure[\ $z-8\mathcal{G}_3 M$
		\label{FIGzM}]{\includegraphics[width=4.3cm]{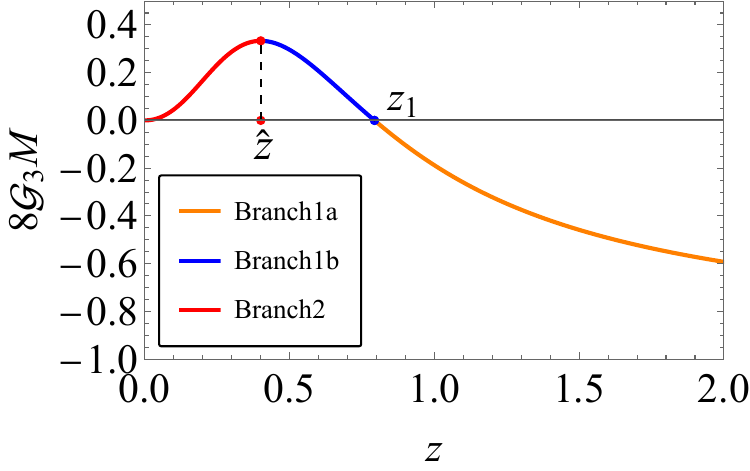}}
		\
		
		\subfigure[ \ $z-T \pi \ell_3 $: $\nu>1$
		\label{FIGzTL1}]{\includegraphics[width=4.3cm]{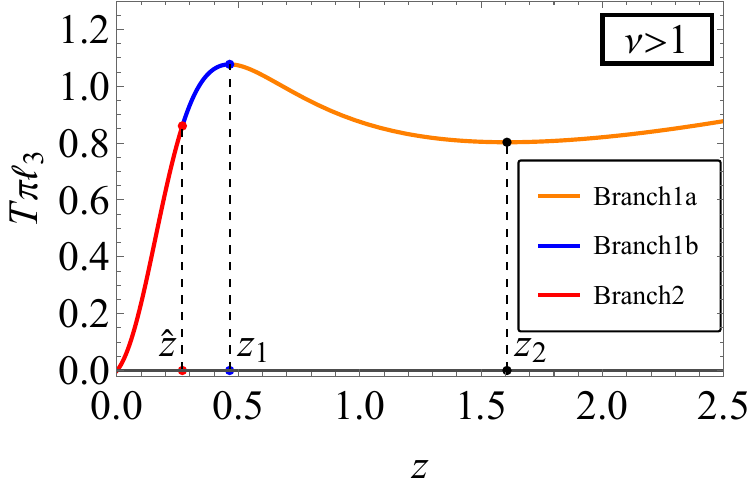}}
		\subfigure[\ $z-T  \pi \ell_3$: $\nu<1$
		\label{FIGzTl1}]{\includegraphics[width=4.3cm]{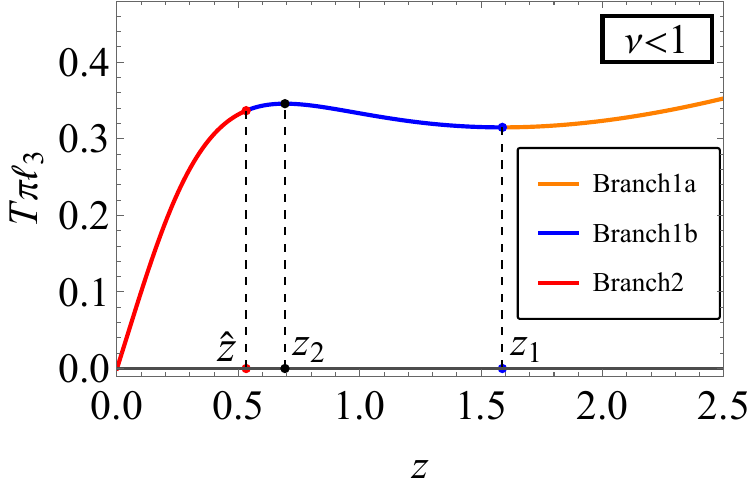}}
	\end{center}
	\caption{Thermodynamical quantities as a function of parameter $z$. Here $z_1$, $z_2$ and $\hat{z}$ are given by Eqs.~\eqref{eq_z1z2} and \eqref{eq_hatz}. We set $\nu=2$ in Fig.~\ref{FIGzS} and \ref{FIGzM}, $\nu = 10$ in Fig.~\ref{FIGzTL1}, and $\nu = 0.25$  in Fig.~\ref{FIGzTl1}.}\label{FIG_z_Thermodynamic}
\end{figure}

Furthermore, in Fig. \ref{FIG_z_Thermodynamic}, we plot the thermodynamical quantities, entropy, mass, and temperature as a function of $z$. Note that $z_1$, $z_2$, and $\hat{z}$ defined by Eqs.~\eqref{eq_z1z2} and \eqref{eq_hatz} are clearly shown. Interestingly, two different quBTZ black holes (with different values of $z$) can possess the same entropy, which distinguishes them from these classical BTZ black holes. Another distinction is that the mass of the quBTZ black hole is constrained in $-1 \leqslant 8 \mathcal{G}_3 M \leqslant 1/3$.

\section{Thermodynamical topology of quantum BTZ black hole}\label{Sec_ThermTopology}

In recent studies, significant attention has been devoted to explore the intriguing properties of the quBTZ black holes. These findings significantly suggest that quBTZ black holes possess a unique phase structure that deviates from the classical BTZ black holes. Consequently, we aim to investigate the potential distinctive features of quBTZ black holes from the perspective of thermodynamical topology.

Following Ref.~\cite{Wei:2022BHasTopDefect}, the generalized free energy of the quBTZ black holes can be defined as
\begin{equation}
	F_g =M - \frac{1}{\tau} S= \frac{z^2(1+\nu z)\left( 1-\nu z^3 \right)}{2\mathcal{G} _3\left( 2\nu z^3+3z^2+1 \right) ^2}-\frac{1}{\tau}\frac{\pi \ell _3z}{\mathcal{G} _3\left( 2\nu z^3+3z^2+1 \right)}. \label{eq_generaliedF}
\end{equation}
The parameter $\tau$ has a dimension of time, and can be interpreted as the inverse of the temperature of the surrounding black hole's environment. Obviously, via this definition, it can be found that when $\tau^{-1}$ is equal to the Hawking temperature \eqref{eq_temperature}, the generalized free energy will be taken to its extreme value and correspond to the on-shell free energy.

To construct the thermodynamical topology, the vector field mapping
\begin{equation}
	\phi : X=\left\{ \left( z,\theta \right) \middle| 0<z<\infty ,0<\theta <\pi \right\} \rightarrow \mathbb{R} ^2, \label{eq_defphi}
\end{equation}
can be defined as
\begin{equation}
	\phi \left(z, \theta \right) = \left( \frac{\partial_{z} F_g}{\partial_{z} S} , -\cot \theta \csc \theta   \right) = \left( \frac{z\left( \nu z^3+3\nu z+2 \right)}{2\pi \ell_3\left( 2 \nu z^3+3z^2+1 \right)}-\frac{1}{\tau},-\cot \theta \csc \theta \right). \label{eq_mapphi}
\end{equation}
As the parameter $z$ characterizes the quBTZ black holes, it becomes the first parameter for the domain of mapping $\phi$. The other parameter, $\theta$, in the domain serves as an auxiliary function and is utilized to construct the second component of the mapping, $\phi^\theta = - \cot \theta \csc\theta$. By examining Eq.~\eqref{eq_mapphi}, it becomes evident that the zero point of $\phi$ corresponds to the black hole with a temperature of $T = \tau^{-1}$ (with $\theta$ taking the value of $\pi/2$). Thus, the zero point of the mapping $\phi$ can characterize the black hole solution with a fixed $\tau$.

Furthermore, utilizing Duan's $\phi$-mapping topological current theory \cite{Duan:1979SU2,Duan:1984Structure}, the zero points of the mapping $\phi$ can be linked to the topological number. Specifically, the topological number can be obtained by calculating the weighted sum of the zero points. The weight assigned to each zero point is determined by its nature. For example, a saddle point has a weight of $-1$ and an extremum point has a weight of $1$. Additionally, this weight can also be evaluated by using the winding number
\begin{equation}
	w_{(z,\theta )}=\frac{1}{2\pi}\int_{C_{\delta}\left( z,\theta \right)}{d\Omega}, \label{eq_windingnumber}
\end{equation}
where $C_{\delta}(z, \theta)$ represents a sufficiently small closed loop that encircles the point $(z, \theta)$, and $d\Omega$ denotes the change in the direction of the vector field $\phi(z, \theta)$ along the curve $C_{\delta}(z, \theta)$. Eventually, this topological number can also be determined by analyzing the boundary behavior, namely,
\begin{equation}
	W\equiv \sum_{(z,\theta )\in X\land \phi (z,\theta )=0}{w_{(z,\theta )}}=\frac{1}{2\pi}\sum_{(z,\theta )\in X\land \phi (z,\theta )=0}{\int_{C_{\delta}\left( z,\theta \right)}{d\Omega}}=\frac{1}{2\pi}\int_{\partial X}{d\Omega},\label{eq_W}
\end{equation}
where $d\Omega$ measures the change in the direction of the vector field along the boundary $\partial X$. This implies that certain characteristics of the zero points can be obtained via the total topological number without requiring detailed information.

\begin{figure}
	\begin{center}
		\subfigure[ Global boundary of $X$.	\label{FIGCircleztheta1}]{\includegraphics[width=5cm]{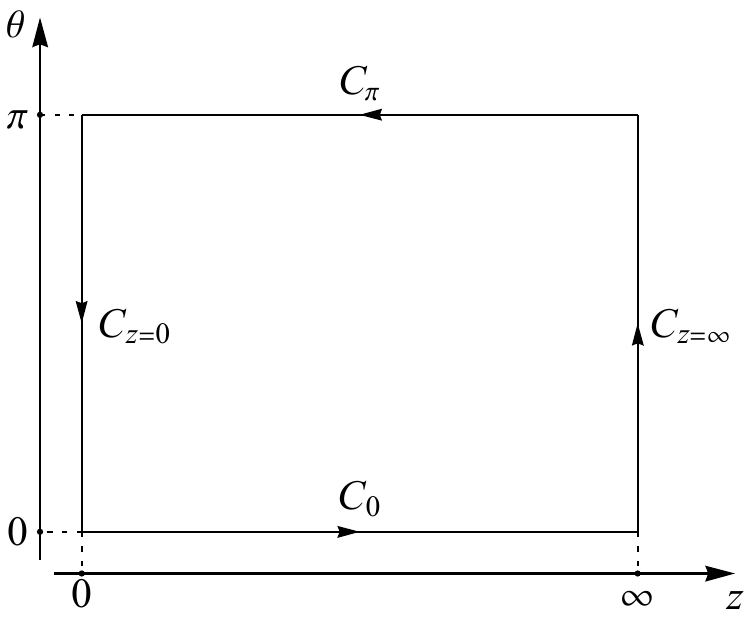}}
		\label{FIG_Circlezthetaglobal}
		\ \ \ \ \ \ \
		\subfigure[ Local boundary of $X_L$ and $X_R$.	\label{FIGCircleztheta2}]{\includegraphics[width=5cm]{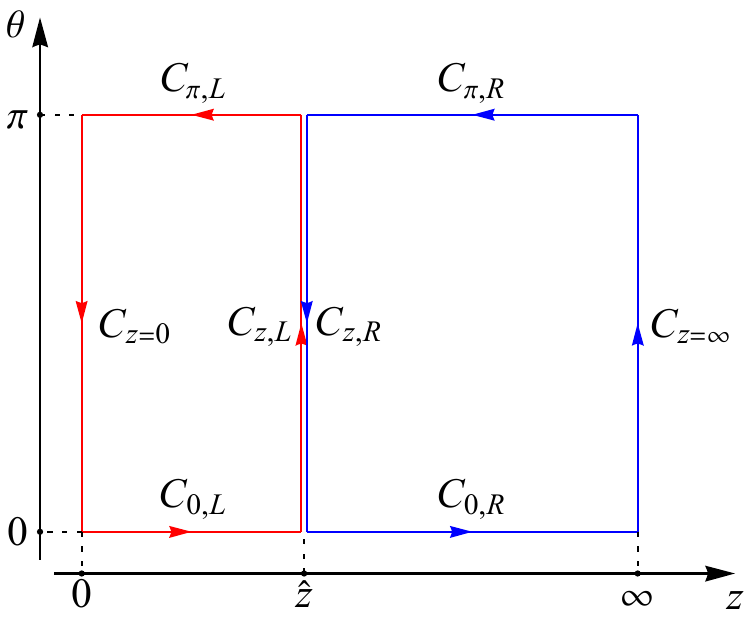}
			\label{FIG_Circlezthetalocal}}
	\end{center}
	\caption{Sketch of the boundary for given domain. (a) represents the boundary of domain $X=\left\{ ( z,\theta ) \middle| 0<z<\infty ,0<\theta <\pi \right\}$. (b) represents the boundaries of regions $X_L=\left\{ \left( z,\theta \right) \middle| 0<z<\hat{z},0<\theta <\pi \right\}$ and $X_R=\left\{ \left( z,\theta \right) \middle| \hat{z}<z<\infty ,0<\theta <\pi \right\}$. }
	\label{FIG_Circleztheta}
\end{figure}

In order to obtain the total topological number of the quBTZ black holes, we sketch these boundaries represented by $C_{z=0}$, $C_{z=\infty}$, $C_0$, and $C_\pi$ in Fig.~\ref{FIGCircleztheta1}. Next, we shall analyze the behavior of the vector at the boundary of domain $X$.

First, for the boundaries $C_0$ and $C_\pi$, we have
\begin{equation}
	\left. \phi ^{\theta} \right|_{\theta \rightarrow 0}=-\frac{1}{\theta ^2}+\mathcal{O}\left( 1 \right) ,\quad \left. \phi ^{\theta} \right|_{\theta \rightarrow \pi}=\frac{1}{(\theta -\pi )^2} +\mathcal{O}\left( 1 \right). \label{eq_boundarytheta}
\end{equation}
Thus along the curves $C_0$ and $C_\pi$, the vector is always downward and upward respectively. Secondly, for the boundaries $C_h$ and $C_\infty$, we have
\begin{equation}
	\left. \phi ^z \right|_{z\rightarrow 0}=-\frac{1}{\tau}+\mathcal{O} \left( z \right) , \left. \phi ^z \right|_{z\rightarrow \infty}=\frac{1}{4\pi \ell _3}z+\mathcal{O} \left( 1 \right),
\end{equation}
which indicates that, along the curves $C_{z=0}$ and $C_{z=\infty}$, the vector is always leftward and rightward respectively. Therefore, based on the behaviors along four segments of the boundary, we can conclude that the topological number
\begin{equation}
	W = 1,	\label{eq_GlobalW}
\end{equation}
which indicates that the difference between the numbers of extremum points and saddle points is equal to $1$.

Let us now turn to the physical interpretation of saddle points and extremum points. From the expression $\phi^{\theta} = - \cot \theta \csc \theta$, one can see that $\partial_\theta \phi^{\theta}|_{\theta = \pi /2} = 1$. It is important to note that the Jacobian determinant of $\phi$ is given by $\partial_z \phi^z \partial_\theta \phi^\theta$. Consequently, extremum points and saddle points correspond to the locations where $\partial_z \phi^z > 0$ and $\partial_z \phi^z < 0$, respectively. Moreover, there exists a relationship between the heat capacity $C$ and $\partial_z \phi^z$,
\begin{equation}
	C=\frac{T}{\partial _z\phi ^z}\partial _z S.
\end{equation}
Consequently, positive and negative values of the heat capacity $C$ can be associated with the extremum points and saddle points, respectively, or with the winding number. After a simple analysis, we show the relation between the thermodynamical stability and winding number in TABLE.~\ref{Table_ThermStabilityWindingN}. Remarkably, the relation relies on the sign of $\partial_z S$, which differs from the discussion given in Ref.~\cite{Wei:2022BHasTopDefect}. On the other hand, when $\partial_z S <0$, the correspondence between the winding number and the stability of the state appears to be anomalous.

\begin{table}
	\centering
	\resizebox{12cm}{!}{
		\begin{tabular}{ccccc}
			\hline
			\ \ \ \ \ \ Case \ \ \ \ \ \ &  \ \ \ \ \ \ Range \ \ \ \ \ \ &  \ \ \ \ \ \ Branch\ \ \ \ \ \  &  \ \ \ \ \ \ Thermodynamical stability\ \ \ \ \ \ & \ \ \ \ \ \ Winding number \ \ \ \ \ \  \\
			\hline
			$ \partial_z S >0 $  & $0<z<\hat{z}$ & Branch 2 & Stable / Unstable & $w = 1$ / $w=-1$  \\
			$ \partial_z S <0 $  & $\hat{z}<z<\infty$ & Branch 1a or 1b & Stable / Unstable  & $w = -1$ / $w=1$ \\
			\hline
	\end{tabular}}
	\caption{The relation between thermodynamical stability and winding number.}
	\label{Table_ThermStabilityWindingN}
\end{table}

\begin{figure}
	\begin{center}
		{\includegraphics[width=6cm]{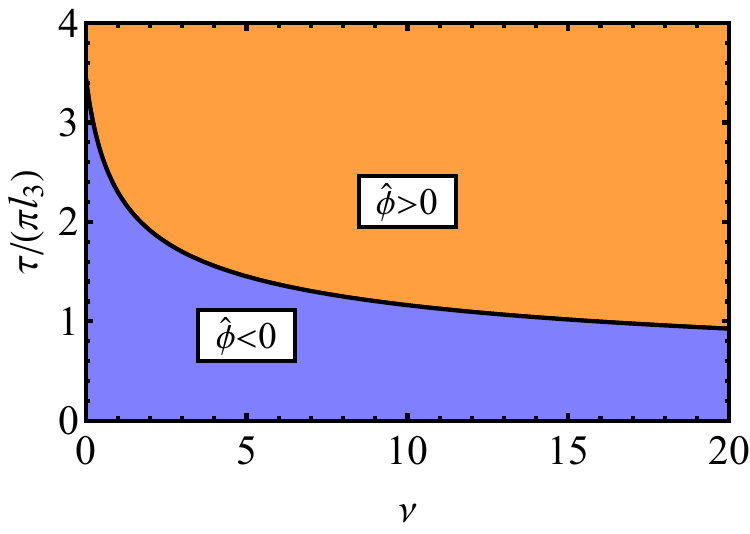}}
	\end{center}
	\caption{Schematic diagram of critical value of $\hat{\phi}$. The blue and orange areas represent the parameter region for $\hat{\phi} < 0$ and $\hat{\phi} > 0$, respectively. The black curve is for $\hat{\phi} = 0$. } 		\label{FIG_nutauphi}
\end{figure}

Hence, to obtain a physical interpretation from the winding number or topological number, the behavior of the vector field on the boundary at $z = \hat{z}$ becomes crucial. This observation motivates us to partition the domain $X$ into two distinct regions, namely,
\begin{align}
	X_L&=\left\{ \left( z,\theta \right) \middle| 0<z<\hat{z},0<\theta <\pi \right\},
	\\
	X_R&=\left\{ \left( z,\theta \right) \middle| \hat{z}<z<\infty ,0<\theta <\pi \right\},
\end{align}
which is shown in Fig.~\ref{FIGCircleztheta2}. The domains $X_R$ and $X_L$ correspond exactly to branch 1 and branch 2. At the boundary $z = \hat{z}$, the vector field reduces to
\begin{equation}
	\left. \phi ^z \right|_{z \rightarrow \hat{z}}= \hat{\phi}+ \frac{1-3\hat{z}^2}{4\pi \ell _3\hat{z}^2}\left( z-\hat{z} \right) + \mathcal{O}\left( (z - \hat{z})^2 \right),
	\label{eq_phizzh}
\end{equation}
wherein,
\begin{equation}
	\hat{\phi} = \frac{1}{\hat{\tau}}-\frac{1}{\tau},\quad \hat{\tau}=\frac{2\pi \ell _3\left( 2\nu \hat{z}^3+3\hat{z}^2+1 \right)}{\hat{z}\left( \nu \hat{z}^3+3\nu \hat{z}+2 \right)},
\end{equation}
and $\hat{z}$ is given by Eq.~\eqref{eq_hatz}. The leading term in Eq.~\eqref{eq_phizzh} is $\hat{\phi}$ unless it vanishes. Therefore, the positive, negative, and zero values of $\hat{\phi}$ hold significance for the local topological number of $X_L$ and $X_R$. It is evident that the sign of $\hat{\phi}$ depends on $\tau/(\pi l_3)$ and $\nu$. This correlation is illustrated in Fig.~\ref{FIG_nutauphi}, where the black curve represents $\tau = \hat{\tau}$.	

The vector behavior at other boundaries $X_L$ and $X_R$ can inherit previous analysis, and the sign of their crucial component remains invariant for any $z>0$ and $\tau>0$. Consequently, the value of $\hat{\phi}$ determines the local topological number $W_L$ for the left region $X_L$ and $W_R$ for the right region $X_R$. In other words, when $\hat{\phi} > 0$, the topological number is
\begin{equation}
	W_L = 1, \quad W_R = 0;
	\label{eq_phi1W}
\end{equation}
and for $\hat{\phi}< 0 $, the topological number reads
\begin{equation}
	W_L = 0, \quad W_R = 1.
	\label{eq_phi2W}
\end{equation}
As indicated by Eqs.~\eqref{eq_W} and \eqref{eq_GlobalW}, the total topological number $W_R + W_L$ is equal to $W$ as expected in both the cases: $\hat{\phi} > 0$ and $\hat{\phi} < 0$. However, it is important to note that the physical interpretations of Eq.~\eqref{eq_phi1W} ($\hat{\phi} > 0$) and Eq.~\eqref{eq_phi2W} ($\hat{\phi} < 0$) are distinct, as demonstrated in TABLE.~\ref{Table_ThermStabilityWindingN}.

For $\hat{\phi} = 0$, we observe that $\phi^z(\hat{z}) = 0$, indicating that the zero point resides at the boundary $z = \hat{z}$. In such cases where the boundary behavior leads to a zero point, the local topological number becomes invalid. However, if we consider the boundary $z = \hat{z}$ to be very close to $z = \hat{z}$, the topological number can still be effectively determined. In Fig.~\ref{FIG_Circlezthetalocal}, $C_{z,L}$ and $C_{z,R}$ should be slightly to the left and right of the boundary $z = \hat{z}$, respectively. Within this framework, for $\hat{\phi} = 0$, the first-order term in Eq.~\eqref{eq_phizzh} vanishes, and the second-order term becomes significant. As $1-3\hat{z}^2$ is always greater than zero, the topological number can still be read from the vector field on the boundary
\begin{equation}
	W_L = 0, \quad W_R = 0.\label{eq_phi0W}
\end{equation}
In this scenario, the sum of $W_L$ and $W_R$ is equal to $0$, which does not match the total topological number. This discrepancy arises because when $\hat{\phi} = 0$, $z = \hat{z}$ also represents a zero point that is not included in the regions $X_L$ and $X_R$. Furthermore, we observe that the zero point $(\hat{z}, \pi/2)$ corresponds to a minimum point, contributing a value of $1$ to the topological number. Thus, despite $W_L + W_R = 0$, the total topological number remains $1$. Note that the case $z = \hat{z}$ represents a black hole solution with $\tau = \hat{\tau}$, characterized by a vanishing heat capacity ($C = 0$). Further discussion regarding the black hole state at $z = \hat{z}$ is provided in Appendix~\ref{A_zeqzhat}.

In summary, considering TABLE.~\ref{Table_ThermStabilityWindingN} and the previous discussion, we can obtain the physical interpretation of the local topological number for the quBTZ black holes. There are three cases that require further examination, excluding the state $z = \hat{z}$. These cases are as follows:
\begin{itemize}
	\item The case of $\hat{\phi}>0$. The result $W_L = 1$ implies that there is one more stable black hole states than the unstable states for branch 2. $W_R = 0$ indicates that the numbers of the stable and unstable states are equal for branch 1. Consequently, in total, we have one more stable black hole state.
	\item The case of $\hat{\phi}<0$. The result $W_L = 0$ implies that the numbers of the stable and unstable black hole states are equal for branch 2. Meanwhile, $W_R = 1$ indicates that there is one more unstable states than the stable states for branch 1.Therefore, in total, we have one more unstable black hole state.
	\item The case of $\hat{\phi} = 0$. The result $W_L = 0$ and $W_R = 0$ indicates that the numbers of the stable and unstable black hole states are equal for either branch 2 or branch 1.
\end{itemize}

Note that the total topological number always remains $1$; however, it no longer reflects the difference in number between the stable and unstable phase states. This suggests that the topological number defined by $\phi$ in Eq. \eqref{eq_defphi} does not adequately describe the thermodynamical properties of the quBTZ black holes. On the other hand, when we consider the boundary $z = \hat{z}$ or the point where $\partial_z S = 0$, the local topological number can accurately correspond to the physical behavior according to TABLE.~\ref{Table_ThermStabilityWindingN}. This sheds light on the underlying physics associated with the mapping of $\phi$. In Sec.~\ref{Sec_Example}, we will provide several illustrative examples to further confirm the aforementioned discussion.

\section{Revisit thermodynamical topology of quantum BTZ black hole}\label{Sec_Revisit}

Based on the preceding discussion, it becomes apparent that there are certain issues associated with the topological number defined by $\phi$ in Eq.~\eqref{eq_defphi}. While the zero points of $\phi$ can indeed represent the black hole solutions, the main concern is that the topological number does not directly relate to the stability of black holes. To address this problem, it is necessary to introduce a local topological number for a more meaningful physical interpretation by taking into account anomalous corresponding regions indicated by $\partial_z S < 0$. In this section, we will reexamine the topological aspects of the black hole thermodynamics and propose an alternative variable, denoted as $\Phi$, to characterize it.

From Fig.~\ref{FIGzS}, one can see that the function between $z$ and $S$ is smooth but not injective, which leads to the existence of regions of $\partial_z S > 0$ and $\partial_z S < 0$. Looking back at the physical meanings of $z$ and $S$, we can see that
\begin{itemize}
	\item The parameter $z$ serves as a characterization of the quBTZ black holes. In light of this, we previously constructed $\phi$ in equation \eqref{eq_defphi} with the domain $X$.
	\item The parameter $S$ represents the thermodynamical entropy and is a natural quantity in thermodynamics. Meanwhile, the heat capacity can be expressed as $T \partial S/ \partial T$.
\end{itemize}
Therefore, to accurately describe the thermodynamical topology of the quBTZ black holes, a new approach would be introduced via employing a vector mapping with the domain defined by the entropy $S$. However, it should be noted that one entropy $S$ can correspond to two different parameters $z$, resulting in a ``one to two" relationship between $S$ and $z$, as depicted in Fig.~\ref{FIG_ThermzS}. This gives rise to two distinct foliations
\begin{align}
	&z = f_1\left(S\right), \quad \text{for } \hat{z} < z < \infty,
	\label{eq_foliationz1}
	\\
	&z = f_2\left(S\right), \quad \text{for } 0<z<\hat{z},
	\label{eq_foliationz2}
\end{align}
with domain $0 < S < \hat{S}$. The explicit expressions of $f_1(S)$ and $f_2(S)$ are given by \eqref{eq_f1(S)} and \eqref{eq_f2(S)}. It is worth noting that the foliation 1 given by Eq.~\eqref{eq_foliationz1} corresponds precisely to branch 1, while foliation 2 described by Eq.~\eqref{eq_foliationz2} corresponds to branch 2. As a result, the generalized free energy given by Eq.~\eqref{eq_generaliedF} can be divided into two distinct parts
\begin{equation}
	{F_g}_1\left( S \right) =\left. F_g \right|_{z=f_1\left( S \right)},\quad {F_g}_2\left( S \right) =\left. F_g \right|_{z=f_2\left( S \right)}.
	\label{eq_SFg}
\end{equation}
Correspondingly, there are two vector mappings
\begin{align}
	\Phi_1 : Y=\left\{ \left( S,\theta \right) \middle| 0<S<\hat{S} ,0<\theta <\pi \right\} \rightarrow \mathbb{R} ^2,
	\label{eq_defPhi1}
	\\
	\Phi_2: Y=\left\{ \left( S,\theta \right) \middle| 0<S<\hat{S} ,0<\theta <\pi \right\} \rightarrow \mathbb{R} ^2,
	\label{eq_defPhi2}
\end{align}
which are given by
\begin{align}
	\Phi _1\left( S,\theta \right) =\left( \partial _S{F_g}_1,-\cot \theta \csc \theta \right) =\phi \left( f_1\left( S \right), \theta \right),
	\label{eq_Phi1def}
	\\
	\Phi _2\left( S,\theta \right) =\left( \partial _S{F_g}_2,-\cot \theta \csc \theta \right) =\phi \left( f_2\left( S \right), \theta \right).
	\label{eq_Phi2def}
\end{align}
For a fixed value of $\tau$, the zero points of the mappings $\Phi_1$ and $\Phi_2$ represent black hole states with a Hawking temperature of $1/\tau$. Furthermore, due to the domain $0 < S < \hat{S}$, the functions $f_1(S)$ and $f_2(S)$ cover the regions $(0,\hat{z}) \cup (\hat{z},\infty)$. Hence, the combination of the zero points for $\Phi_1$ and $\Phi_2$ corresponds to the black hole states with a temperature of $1/\tau$ individually, except for the state with $z = \hat{z}$. The reason for this ``exception" is that the Jacobian determinant of the function $S(z)$ becomes degenerate at $z = \hat{z}$. Consequently, the black hole state $z = \hat{z}$ needs to be further considered, and its discussion can be found in Appendix~\ref{A_zeqzhat}.

Moreover, by Eqs.~\eqref{eq_Phi1def} and \eqref{eq_Phi2def}, the heat capacity can be given by
\begin{equation}
	C=\frac{T}{\partial _S{\Phi _1^S}},\quad C=\frac{T}{\partial _S{\Phi _2^S}}.
	\label{eq_CSPhi}
\end{equation}
Therefore, through a similar analysis, we can establish the relationship between the thermodynamical stability and the winding number. Specifically, the term ``stable" (``unstable") corresponds to a winding number of $w = 1$ ($w = -1$), which is consistent with the findings of Ref.~\cite{Wei:2022BHasTopDefect}. This property is completely different from the vector field defined by Eqs.~\eqref{eq_defphi} and \eqref{eq_mapphi}, as demonstrated in TABLE.~\ref{Table_ThermStabilityWindingN}. Furthermore, Eq.~\eqref{eq_CSPhi} indicates that $\Phi$ ($\Phi_1$ and $\Phi_2$) is more nature to characterize the thermodynamical topology of quBTZ black holes.

\begin{figure}
	\begin{center}
		{\includegraphics[width=4.5cm]{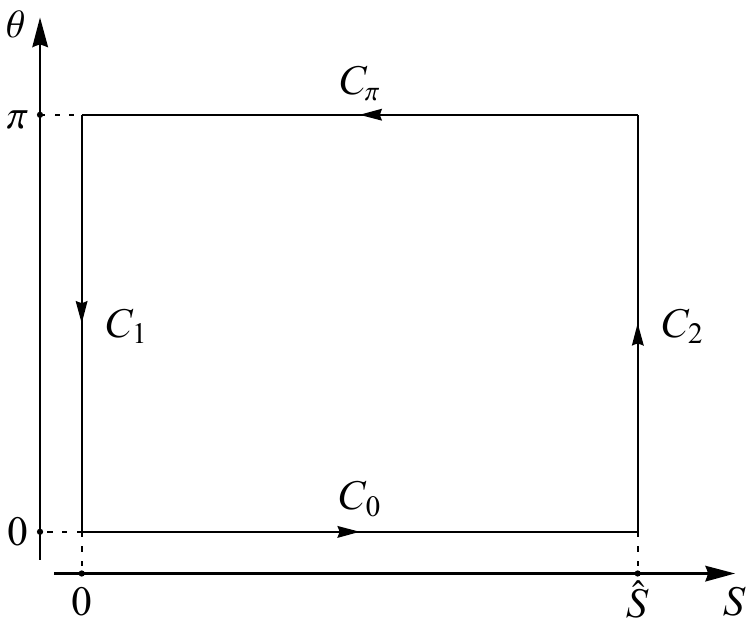}}
	\end{center}
	\caption{Sketch of boundary of region $Y = \left\{ (S, \theta) \middle| 0<S<\hat{S} ,0<\theta <\pi \right\} $.}
	\label{FIG_CircleStheta}
\end{figure}

Based on the previous discussion, there are two foliations that require further investigation. The outline of the boundary has been depicted in Fig.~\ref{FIG_CircleStheta}. Additionally, the values of the vector fields $\Phi_1(S,\theta)$ and $\Phi_2(S,\theta)$ at two crucial boundaries of $Y$ are as follows:
\begin{itemize}
	\item Foliation 1:
	For the boundaries $S = 0$ and $S = \hat{S}$, we have
	\begin{align}
		\left. \Phi _{1}^{S} \right|_{S\rightarrow 0}&=\frac{1}{4\sqrt{2\pi \ell \mathcal{G} _3S}}+\mathcal{O} \left( 1 \right),\\
		\left. \Phi _{1}^{S} \right|_{S\rightarrow \hat{S}} &=\frac{1}{\hat{\tau}}-\frac{1}{\tau} + \frac{1-3\hat{z}^2}{4\pi \ell _3\hat{z}^2}\sqrt{\frac{3\hat{z}\left( \hat{z}^2+1 \right) ^2}{2\left( 1-\hat{z}^2 \right)}\frac{\mathcal{G} _3}{\pi \ell _3}(\hat{S}-S)} + \mathcal{O} \left( (S - \hat{S}) \right)
		\\
		&= \hat{\phi} +\frac{1-3\hat{z}^2}{4\pi \ell _3 \hat{z}^2}\sqrt{\frac{3\hat{z}\left( \hat{z}^2+1 \right) ^2}{2\left( 1-\hat{z}^2 \right)}\frac{\mathcal{G} _3}{\pi \ell _3}(\hat{S}-S)} + \mathcal{O} \left( (S - \hat{S}) \right) .
		\label{eq_Phi1ShatS}
	\end{align}
	\item Foliation 2:
	For the boundaries $S = 0$ and $S = \hat{S}$, we have
	\begin{align}
		\left. \Phi _{2}^{S} \right|_{S\rightarrow 0}&=-\frac{1}{\tau}+\mathcal{O} \left( S \right),\\
		\left. \Phi _{2}^{S} \right|_{S\rightarrow \hat{S}} &=\frac{1}{\hat{\tau}}-\frac{1}{\tau} - \frac{1-3\hat{z}^2}{4\pi \ell _3\hat{z}^2}\sqrt{\frac{3\hat{z}\left( \hat{z}^2+1 \right) ^2}{2\left( 1-\hat{z}^2 \right)}\frac{\mathcal{G} _3}{\pi \ell _3}(\hat{S}-S)} + \mathcal{O} \left( (S - \hat{S}) \right)
		\\
		&= \hat{\phi} -\frac{1-3\hat{z}^2}{4\pi \ell _3 \hat{z}^2}\sqrt{\frac{3\hat{z}\left( \hat{z}^2+1 \right) ^2}{2\left( 1-\hat{z}^2 \right)}\frac{\mathcal{G} _3}{\pi \ell _3}(\hat{S}-S)} + \mathcal{O} \left( (S - \hat{S}) \right).
		\label{eq_Phi2ShatS}
	\end{align}
\end{itemize}
On the other hand, the boundaries $\theta = 0$ and $\theta = \pi$ are also given by Eq.~\eqref{eq_boundarytheta}. As a result, it is not difficult to get the topological number, which depends on the sign of $\hat{\phi}$. For $\hat{\phi} > 0$, the topological number
\begin{equation}
	W_1 = 0, \quad W_2 = 1, \quad W = W_1+W_2 = 1;
	\label{eq_Phi1W}
\end{equation}
and for $\hat{\phi} < 0$, the topological number
\begin{equation}
	W_1 = -1, \quad W_2 = 0, \quad W = W_1+W_2 = -1.
	\label{eq_Phi2W}
\end{equation}
For $\hat{\phi}= 0$, when setting the boundary as the line $S = \hat{S}$, the topological number becomes invalid since $\Phi_1^{S}(\hat{S})= 0$ and $\Phi_2^{S}(\hat{S})= 0$. Similar to the discussion given in Sec.~\ref{Sec_ThermTopology}, it is necessary to define the boundary as a line very close to $S = \hat{S}$. Moreover, due to $\hat{\phi} = 0$, the second-order term in Eq.~\eqref{eq_Phi1ShatS} or \eqref{eq_Phi2ShatS} becomes the leading term, resulting in the topological number
\begin{equation}
	W_1 = 0, \quad W_2 = 0, \quad W = W_1+W_2 = 0.
\end{equation}
It is important to note that $S = \hat{S}$ corresponds to $z = \hat{z}$, which falls outside the domain $Y$ covered by the mappings $f_1$ and $f_2$. As a result, this topological number is applicable only to the region $(0,\hat{z}) \cup (\hat{z},\infty)$.

Another noteworthy point is the presence of a phase transition in the topological number determined by $\Phi$, which differs from the topological number determined by $\phi$. Specifically, when $\hat{\phi} > 0$ is satisfied, Eq.~\eqref{eq_Phi1W} yields $W = 1$, while Eq.~\eqref{eq_phi1W} also yields $W = 1$. On the other hand, when $\hat{\phi} < 0$ holds, Eq.~\eqref{eq_Phi2W} gives $W = -1$, while Eq.~\eqref{eq_phi2W} still gives $W = 1$. As discussed in Sec.~\ref{Sec_ThermTopology}, the topological number determined by $\phi$ lacks a direct physical interpretation, requiring that the vector field on the boundary $z = \hat{z}$ provides a meaningful explanation. In contrast, the topological number determined by $\Phi$ is physically meaningful, representing the difference between the number of stable and unstable black hole states. Moreover, this topological transition can be linked to the interpretations provided by the local topological numbers $W_L$ and $W_R$ (determined by $\phi$). We will further explore this through examples given in Sec.~\ref{Sec_Example}.

\section{Some representative examples for thermodynamical topology}\label{Sec_Example}

In Sects.~\ref{Sec_ThermTopology} and \ref{Sec_Revisit}, we have explored the thermodynamical topology characterized by $\phi$ and $\Phi$, respectively. Notably, we have observed a remarkable distinction between the topological numbers defined by $\Phi$ and $\phi$. While the topological number defined by $\Phi$ exhibits a topological phase transition, the one defined by $\phi$ does not. This reason arises from the fact that the topological number defined by $\phi$ lacks certain physical significance. In order to provide a physical interpretation, it becomes necessary to introduce the local topological numbers, denoted as $W_L$ and $W_R$, for the regions on the left and right sides of the boundary $z = \hat{z}$, respectively. Importantly, the physical explanations for $W_L$ and $W_R$ differ from each other. On the contrary, the topology determined by $\Phi$ is physical meaningful. This comes from the involvement of two vector fields, $\Phi_1$ and $\Phi_2$, with the topological number being the sum of $W_1$ and $W_2$.

In this section, we will illustrate the physical implications underlying in $\phi$ and $\Phi$ via two representative examples. To simplify the analysis, we rescale $S$ and $\tau$ by setting $\ell_3 = 1/\pi$ and $\mathcal{G}_3 = 1$. Furthermore, we consider $\nu=10$ as a specific case. In this scenario, the critical condition $\hat{\phi} = 0$ corresponds to $\tau = \hat{\tau} \approx 1.162$. Referring to Fig.~\ref{FIG_nutauphi}, we observe that $\tau > \hat{\tau}$ and $\tau < \hat{\tau}$ correspond to $\hat{\phi} > 0$ and $\hat{\phi} < 0$, respectively. Based on this observation, we examine the cases $\tau = 1.24$ and $\tau = 1$ for further discussion.

Considering $\nu = 10$ and $\tau = 1.24$, we observe that $\hat{\phi}$ is greater than zero, which implies that:
\begin{itemize}
	\item From the mapping of $\phi$, we observe that the topological number is $1$, whereas the local topological numbers yield $W_L = 1$ and $W_R = 0$. This implies that at least one stable black hole state exists within the range of $0 < z < \hat{z}$, while the stable and unstable black hole states appear in pairs within the range of $\hat{z} < z < \infty$.
	\item From the mapping of $\Phi$, we find that the topological number is $1$, which contributed by $W_2 = 1$ from foliation 2 and $W_1 = 0$ from foliation 1. This implies the existence of at least one stable black hole state.
\end{itemize}

\begin{figure}
	\centering
	\includegraphics[width=5 cm]{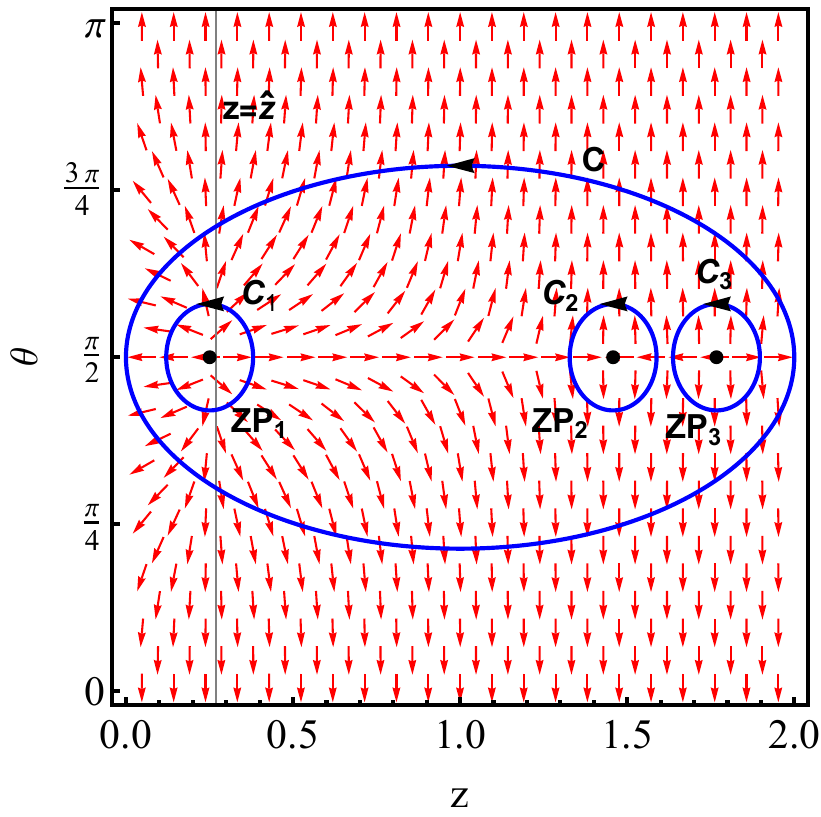}
	\caption{ The diagram of vector field constructed by $\phi$ with $\ell_3 = 1/\pi$, $\mathcal{G}_3 = 1$, $\nu = 10$, and $\tau = 1.24$. The zero points ($ZP_s$) marked with black dots are approximately at $(z,\theta) = (0.25,\pi/2), (1.46,\pi/2)$ and $(1.77,\pi/2)$ for $ZP_1$, $ZP_2$ and $ZP_3$, respectively. This light gray thin line represents the boundary $z=\hat{z} \approx 0.27$. The blue contours $C_i$ are closed loops enclosing the zero points $ZP_i$, and the contours $C$ enclose all zero points. Moreover, the heat capacity for each zero points are $C_{ZP_1} \approx 0.021, C_{ZP_2} \approx  0.47$ and $C_{ZP_3} \approx  -0.32$, respectively. }
	\label{FIG_phiz124}
\end{figure}

\begin{figure}
	\centering
	\includegraphics[width=5 cm]{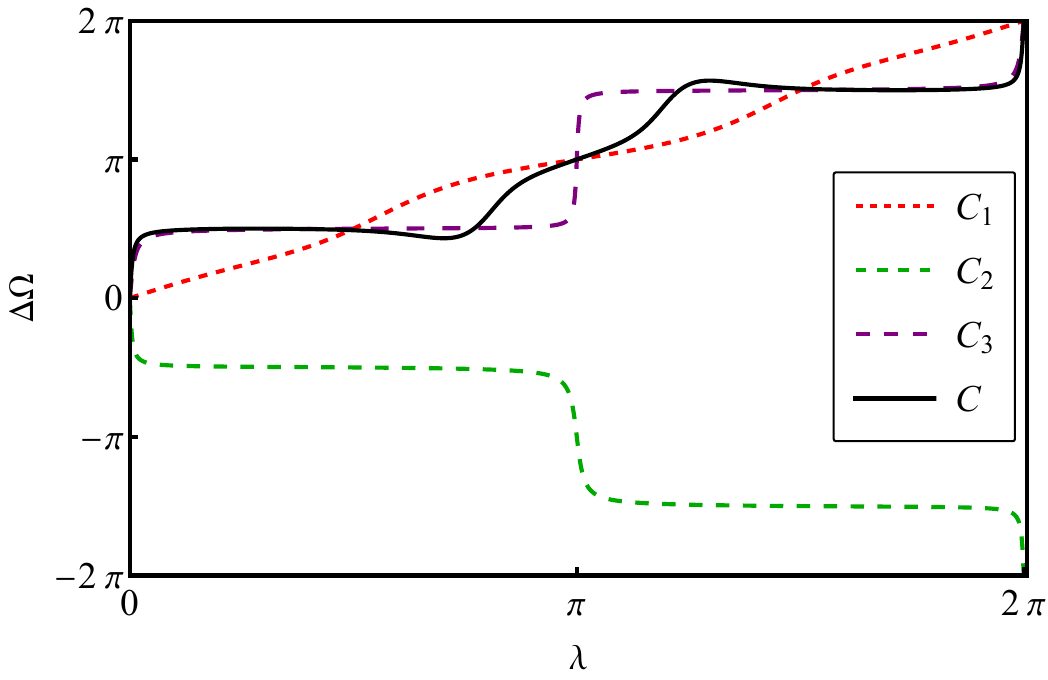}
	\caption{ The curve integral of vector along the different contours. And the winding number can be given by Eq.~\eqref{eq_windingnumber}. These contours integrals correspond to the curve in Fig.~\ref{FIG_phiz124}. Here, $\lambda$ is the angle parameter of the curve with respective to corresponding black point in Fig.~\ref{FIG_phiz124}, and $\Delta\Omega$ is the angle change of the vector field $\phi$ along this curve. Obviously, the winding number of each zero points are given by $w_{ZP_1} = 1, w_{ZP_2} = -1$ and $w_{ZP_3} = 1$.}
	\label{FIG_phiz124Curve}
\end{figure}

Using the specified parameter values, we show the vector field pattern described by Eq.~\eqref{eq_mapphi} in Fig.~\ref{FIG_phiz124}. To focus on the relevant points with non-zero winding numbers, we only present the range of $z$ as $0<z<2$. Notably, we observe three key points: $ZP_1$, an extreme point with a winding number of $w = 1$; $ZP_2$, a saddle point with $w = -1$; and $ZP_3$, another extreme point with $w = 1$. These points signify the relation between the saddle points or extreme points and their corresponding winding numbers. Moreover, by performing the contour integral, we illustrated the results in Fig.~\ref{FIG_phiz124Curve}.


The zero point $ZP_1$ belongs to branch 2, whereas $ZP_2$ and $ZP_3$ are zero points in branch 1. Referring to TABLE.~\ref{Table_ThermStabilityWindingN}, we observe that $ZP_1$ and $ZP_2$ are stable, while $ZP_3$ is unstable. It may initially appear peculiar that black hole solutions with adjacent $z$ parameters, such as $ZP_1$ and $ZP_2$, are both stable. However, this discrepancy can be attributed to the fact that $ZP_1$ and $ZP_2$ reside in different branches, where the physical interpretations of zero points (saddle point or extreme point) differ. This observation confirms the significant dependence of the physical interpretation of the topological number provided by $\phi$ on the specific region we considered.

\begin{figure}
	\begin{center}
		\subfigure[ Foliation 2: The vector field $\Phi_2$ with $0<S<\hat{S}$.	\label{FIG_Phi2S124}]{\includegraphics[width=5cm]{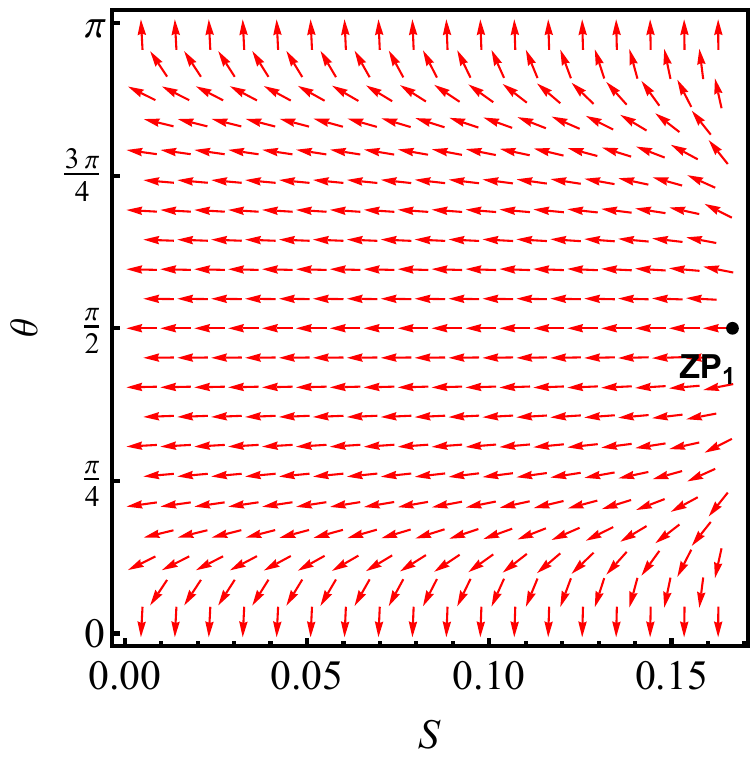}}
		\ \ \
		\subfigure[ Foliation 2: The vector field $\Phi_2$ with $0.1660<S<\hat{S}$. 	\label{FIG_Phi2S124L}]{\includegraphics[width=5cm]{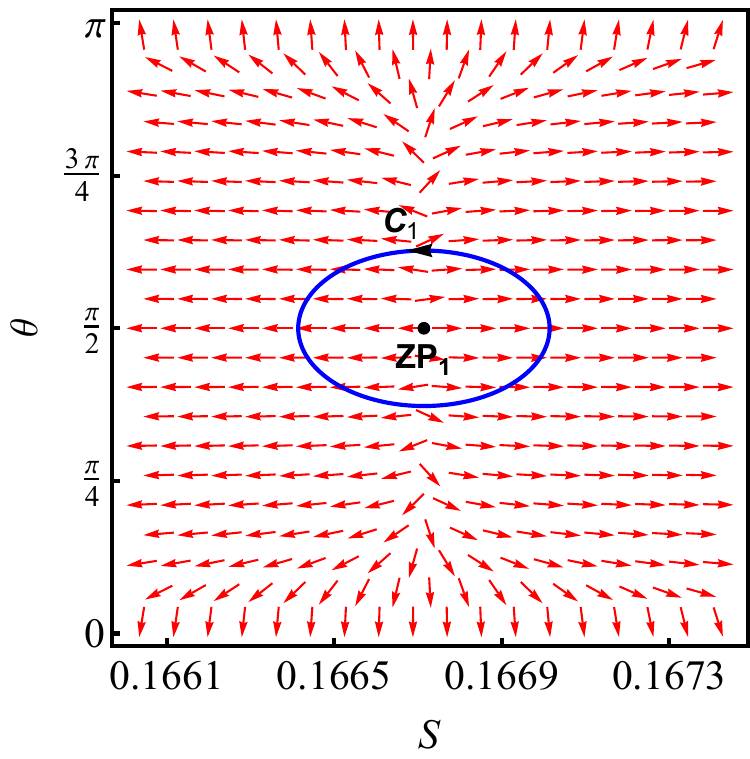}}
		\ \ \
		
		\subfigure[ Foliation 1: The vector field $\Phi_1$ with $0<S<\hat{S}$. \label{FIG_Phi1S124}]{\includegraphics[width=5cm]{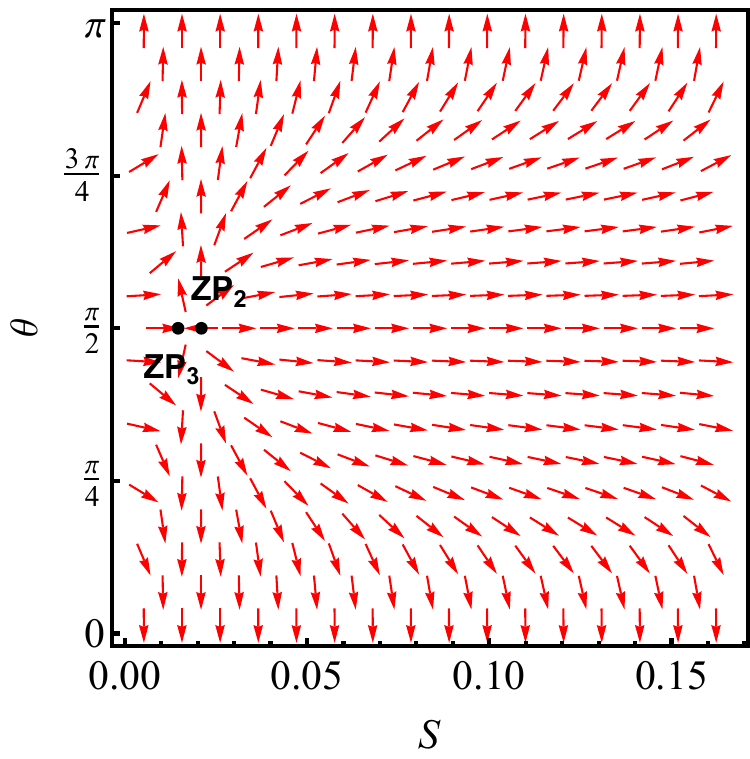}}
		\ \ \
		\subfigure[ Foliation 1: The vector field $\Phi_1$ with $0<S<0.035$. 	\label{FIG_Phi1S124L}]{\includegraphics[width = 5cm]{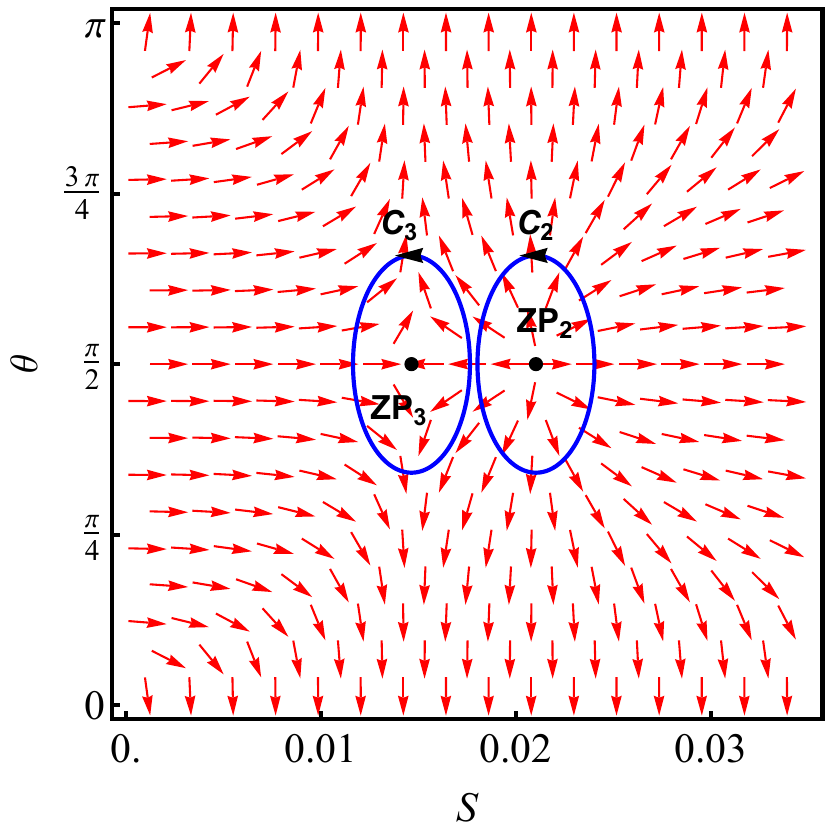}}
	\end{center}
	\caption{The diagram of vector field constructed by $\Phi$ with $\ell_3 = 1/\pi$, $\mathcal{G}_3 = 1$, $\nu = 10$, and $\tau = 1.24$. (b) and (d) are enlarged views of (a) and (c), respectively. The zero points marked with black dots are approximately at $(S,\theta) = (0.1667,\pi/2)$, $(0.0210,\pi/2)$, and $(0.0146,\pi/2)$ for $ZP_1$, $ZP_2$ and $ZP_3$, respectively. These points correspond to the points in Fig.~\ref{FIG_phiz124}.}
	\label{FIG_Phi124}
\end{figure}

On the contrary, when considering $\Phi$ instead of $\phi$, we observe the same physical interpretations of the zero points (saddle point or extreme point) for both branches 1 and 2, as discussed in Sec.~\ref{Sec_Revisit}. The quantity $\Phi$ is described by two foliations, namely $\Phi_1$ and $\Phi_2$. Consequently, we need to consider two vector fields, as illustrated in Fig.~\ref{FIG_Phi124}. In the figure, $ZP_1$, $ZP_2$, and $ZP_3$ correspond to the zero points. Obviously, from the topology constructed by $\Phi$, the winding numbers of $ZP_1$, $ZP_2$, and $ZP_3$ are $1$, $1$, and $-1$, respectively, with the corresponding stability being stable, stable, and unstable. Therefore, the topology given by $\Phi$ consistently provides the same physical interpretation, with the correspondence between the winding number and stability being independent of the branch.

For $\nu = 10$ and $\tau = 1$, we show the vector field $\phi$ and $\Phi$ in Figs. \ref{FIG_phiz100} and \ref{FIG_Phi100} in Appendix \ref{A_Phiphi100}. Similar discussions can be carried out based on these figures, allowing us to verify the conclusions presented in Sec.~\ref{Sec_ThermTopology} and Sec.~\ref{Sec_Revisit}.

\begin{figure}
	\begin{center}
		\subfigure[ \ $z - \tau $	\label{FIG_ztua}]{\includegraphics[width=5cm]{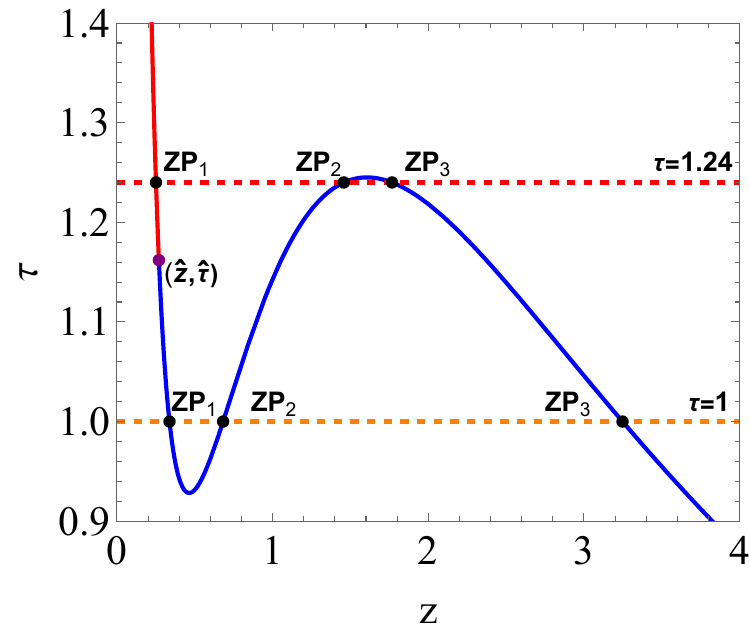}}
		\ \ \ \ \
		\subfigure[ \	$S - \tau $	\label{FIG_Stua}]{\includegraphics[width=5cm]{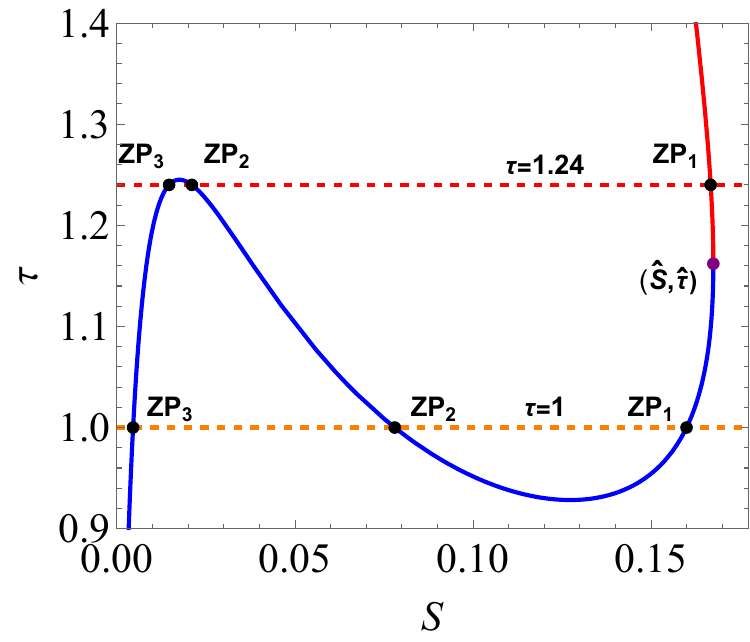}}
	\end{center}
	\caption{(a) $z$ vs. $\tau$. (b) $S$ vs. $\tau$. The parameters take values of $\ell_3 = 1/\pi$, $\mathcal{G}_3 = 1$, and $\nu = 10$. The curve is divided into two colored parts: the red part is for branch 2; the blue part is for branch 1. For $\tau = 1.24$ ($\tau = 1$), $ZP_1$, $ZP_2$ and $ZP_3$ are the corresponding zero points shown in Fig.~\ref{FIG_phiz124} (Fig.~\ref{FIG_phiz100}). Moreover, comparing TABLE.~\ref{Table_ThermodynamicState} and Fig.~\ref{FIG_z_Thermodynamic}, $ZP_1$, $ZP_2$, and $ZP_3$ are for the ``cold" , ``intermediate", and ``hot" black hole, respectively.}
	\label{FIG_zStua}
\end{figure}

One crucial point that deserves emphasis is the transition in the topological number provided by $\Phi$ when $\hat{\phi} > 0$ changes to $\hat{\phi} < 0$. In the case of $\nu = 10$, this transition occurs at the critical point $\tau = \hat{\tau} \approx 1.16$. Specifically, changing $\tau = 1.24$ to $\tau=1$, the topological number changes from $1$ to $-1$. In contrast, the topological number given by $\phi$ remains unchanged. To provide a more physical explanation of this phenomenon, we can plot the curve $\phi^z = 0$ with variables $\tau$ and $z$ in Fig.~\ref{FIG_ztua}, and the curves $\Phi_1^S(S) = 0$ and $\Phi_2^S(S) = 0$ with variables $\tau$ and $S$ in Fig.~\ref{FIG_Stua}. Notably, an interesting observation can be found: for branch 1, the slopes of the respective curves at corresponding points have opposite signs, whereas for branch 2, the slopes have the same sign. Since the thermodynamical stability is determined by the relationship between entropy and temperature, Fig.~\ref{FIG_Stua} directly represents thermodynamical properties. Thus, this anomalous behavior of the topological number in branch 1 can be understood in this context. Furthermore, it can be observed that with $\tau = 1.24$ transitioning to $\tau = 1$, the zero point $ZP_1$ moves from branch 2 to branch 1 via the point $(\hat{z},\hat{\tau})$ or $(\hat{S},\hat{\tau})$, indicating a transition in the topological number.

\section{Discussion and conclusion}\label{Sec_Disscusion}

In this paper, we studied the thermodynamical topology of the quBTZ black holes. Two topologies, defined by $\phi$ (Eq.~\eqref{eq_defphi}) and $\Phi$ (Eqs.~\eqref{eq_defPhi1} and \eqref{eq_defPhi2}) were considered. The former represents a global topological number, but its physical interpretation is region-dependent, as demonstrated in TABLE.~\ref{Table_ThermStabilityWindingN}. On the other hand, the latter provides a more physical and natural global topological number, which captures the difference in the number of stable and unstable black hole phase states, excluding the black hole state $z = \hat{z}$. This topological number plays a similar role to the one in Ref.~\cite{Wei:2022BHasTopDefect}. In fact, the discrepancy between the former and the latter arises from the non-monotonic relationship between the black hole parameter $z$ and entropy $S$. Furthermore, due to the degeneracy in the Jacobian determinant for the mapping from $z$ to $S$ at $z = \hat{z}$, the stability of the phase state $z = \hat{z}$ depends on the branches, as shown in Appendix.~\ref{A_zeqzhat}.

As discussed in Ref.~\cite{Frassino:2023RPTQBTZ}, the quBTZ black holes exhibit three phase states: the ``cold," ``intermediate," and ``hot" black holes. The ``intermediate" black hole is always stable, while the ``hot" black hole is always unstable. However, the ``cold" black hole can exist in both branch 1 and branch 2, with the corresponding stability being unstable and stable, respectively. Consequently, when $\tau > \hat{\tau}$, there is one more stable black hole state than the unstable state, whereas when $\tau < \hat{\tau}$, there is one less stable black hole state. This phenomenon corresponds to the transition of the topological number given by $\Phi$ from $1$ to $-1$, as discussed in Secs.~\ref{Sec_Revisit} and \ref{Sec_Example}. On the other hand, as previously mentioned, the topological number given by $\phi$ remains unchanged. Although this topological number does not easily to understand, we can still obtain the appropriate physical interpretation from the local topological number and its correspondence given in TABLE.~\ref{Table_ThermStabilityWindingN}. In fact, the local topological number given by $\phi$ and the global topological number given by $\Phi$ share the same meaning, focusing on the behavior of the vector field at the boundary defined by $\partial_z {S} = 0$. However, the physical interpretation of the latter is more evident and unified.

On the other hand, the topological number of quBTZ black hole can be $1$ and $-1$, whereas BTZ black holes have a fixed topological number of $1$~\cite{Du:2023BTZ}. When $\tau/(\pi l_3)$ is small, the topological numbers of both are different, and they shall be divided into the different topological classes~\cite{Wei:2022BHasTopDefect}. The topological difference implies that the quantum effects to thermodynamics is significant. On the contrary, when $\tau/(\pi l_3)$ is large, the topological numbers of two black holes are same. And, in this case, the thermodynamic relation $T \sim \mathcal{G} _3S/( \pi l_3 ) ^2$ for quBTZ black hole is nearly identical to those of BTZ black hole. These suggest that at lower temperatures, the influence of quantum correction on thermodynamics is small. Thus, roughly speaking, the variation of strength for quantum effects can correspond to the transition of thermodynamical topological number.

Our results strongly support the fact that when defining topological numbers via the generalized free energy, entropy serves as the most natural choice for the domain variable, establishing a robust physical correspondence. Alternatively, we can select other variables that are monotonically related to $S$ as the domain variable. However, in the cases where there is no one-to-one function between the parameter characterizing black holes and entropy, it becomes necessary to construct foliations for different parameter ranges. This allows for the construction of a one-to-one smooth function between the parameter and entropy within each foliation. The topological number then becomes the sum of contributions from each foliation, with each foliation potentially contributing the numbers $\pm1$ or $0$. This suggests that the thermodynamical topological number may extend beyond the values of $\pm 1$ and $0$.

In summary, our study provides a powerful method for understanding the thermodynamics of quBTZ black holes from a topological perspective, and successfully obtains their thermodynamical topological properties. More detailed properties and features are wished to be disclosed in further study.

\acknowledgments
We are grateful to Prof. Yu-Xiao Liu, Dr. Hong-Yue Jiang, and Dr. Si-Jiang Yang for useful discussions on the quBTZ black holes. This work was supported by the National Natural Science Foundation of China (Grants No. 12075103, No. 12247101).

\appendix
\section{The inverse of function $S = S(z)$}\label{A_z1Sz2S}

For the vector mapping $\Phi$: $Y \rightarrow \mathbb{R}^2 $, the inverse relation between $S$ and $z$ are given by Eq.~\eqref{eq_entropy}, and can be expressed as following two parts,
\begin{align}
	f_1\left( S \right) =\sqrt{\frac{1}{\nu ^2}+\frac{2}{3\nu x}}\sin \left( 	\frac{1}{3}\beta +\frac{\pi}{6} \right) -\frac{1}{2\nu}
	\label{eq_f1(S)},
	\\
	f_2\left( S \right) =\sqrt{\frac{1}{\nu ^2}+\frac{2}{3\nu x}}\cos \left( 	\frac{1}{3}\beta +\frac{\pi}{3} \right) -\frac{1}{2\nu}
	\label{eq_f2(S)},
\end{align}
where
\begin{equation}
	\beta =\cos ^{-1}\left( \frac{3\sqrt{3x}\left( 2x\nu ^2+x+\nu \right)}{(3x+2\nu )^{3/2}} \right) ,\quad x=\frac{\mathcal{G} _3S}{\pi \ell _3}.
\end{equation}
In Fig.~\ref{FIG_ThermzS}, $f_1(S)$ and $f_2(S)$ represent branch 1 and branch 2, respectively. This explicit expression is useful for the discussions presented in our paper.

\section{The black hole solution of $z = \hat{z}$}\label{A_zeqzhat}

\begin{figure}
	\centering
	\includegraphics[width=4 cm]{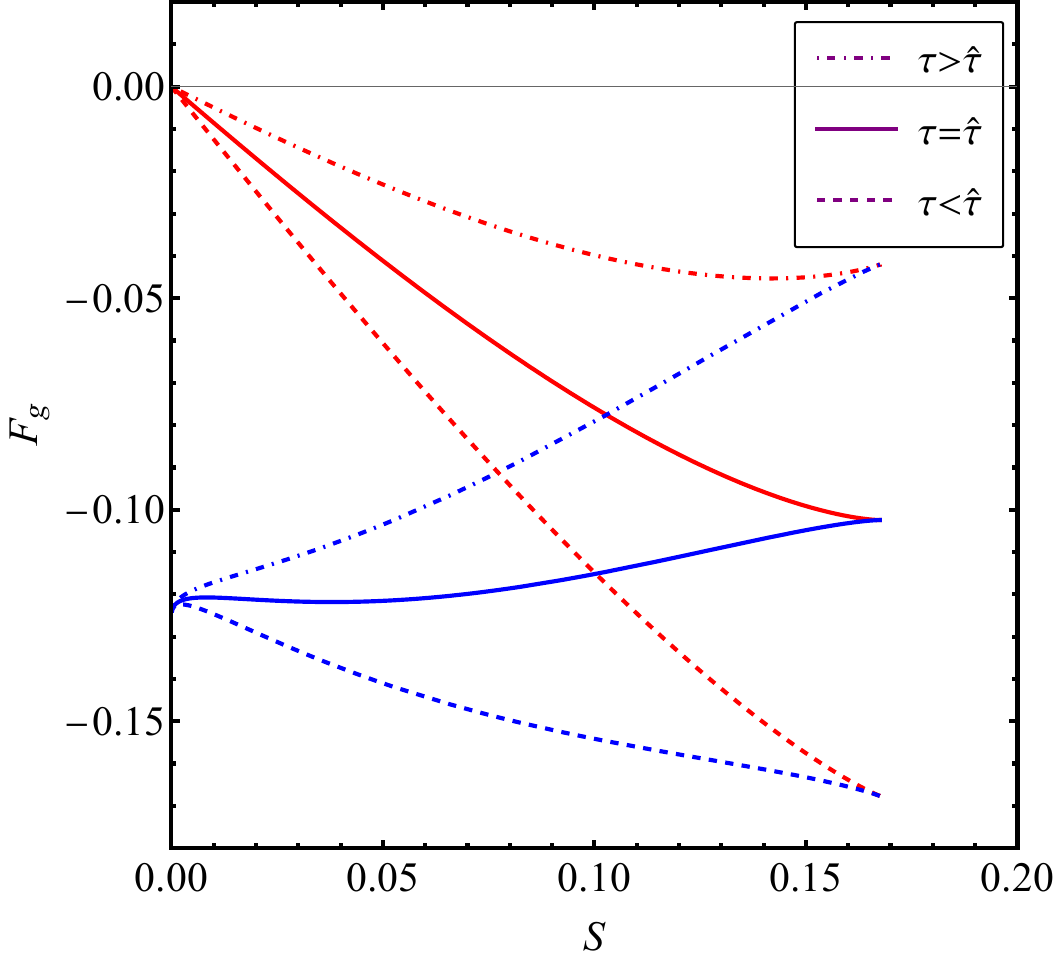}
	\caption{The generalized free energy as a function of entropy for different $\tau$ with $\ell_3 = 1/\pi$, $\mathcal{G}_3 = 1$, and $\nu = 10$. There are three curves for branch 1 (blue) and branch 2 (red). The cases $\tau = \hat{\tau}  \approx 1.162$, $\tau = 2$, and $\tau = 0.8$ are described by solid curves, dot-dashed curves, and and dashed curves, respectively.}
	\label{FIG_SFg}
\end{figure}

As mentioned earlier, it is necessary to separately analyze the stability of the quBTZ black hole phase given by $z = \hat{z}$. The main reason is that this black hole exhibits a zero heat capacity when $\ell_3$, $\nu$, and $\mathcal{G}_3$ remain constant. To address this stability issue, we expand the generalized free energy around $S = \hat{S}$
\begin{align}
	F_{g1}=\hat{F}+\left( \frac{1}{\hat{\tau}}-\frac{1}{\tau} \right) (S-\hat{S})-\frac{1-3\hat{z}^2}{4\pi \ell _3\hat{z}^2}\sqrt{\frac{2\hat{z}\left( \hat{z}^2+1 \right) ^2}{3\left( 1-\hat{z}^2 \right)}\frac{\mathcal{G} _3}{\pi \ell _3}}(\hat{S}-S)^{\frac{3}{2}}+\mathcal{O} \left( (S-\hat{S})^2 \right),
	\label{eq_Fg1Asy}
	\\
	F_{g2}=\hat{F}+\left( \frac{1}{\hat{\tau}}-\frac{1}{\tau} \right) (S-\hat{S})+\frac{1-3\hat{z}^2}{4\pi \ell _3\hat{z}^2}\sqrt{\frac{2\hat{z}\left( \hat{z}^2+1 \right) ^2}{3\left( 1-\hat{z}^2 \right)}\frac{\mathcal{G} _3}{\pi \ell _3}}(\hat{S}-S)^{\frac{3}{2}}+\mathcal{O} \left( (S-\hat{S})^2 \right),
	\label{eq_Fg2Asy}
\end{align}
where $\hat{F} = F_g (\hat{z})$ is given in Eq.~\eqref{eq_generaliedF}. It is important to note that $S$ has an upper bound, denoted as $\hat{S}$, and the asymptotic behavior of $F_{g1}$ and $F_{g2}$ on this boundary is described by Eqs.~\eqref{eq_Fg1Asy} and \eqref{eq_Fg2Asy}. For the case of $\tau > \hat{\tau}$, $S = \hat{S}$ represents a local maximum point for both branches. Conversely, for the case of $\tau < \hat{\tau}$, $S = \hat{S}$ corresponds to a local minimum point for both branches. In the scenario of $\tau = \hat{\tau}$, $S = \hat{S}$ serves as a local minimum point for branch 2 and a local maximum point for branch 1. The corresponding curves are illustrated in Fig.~\ref{FIG_SFg}. If we consider the extreme points of the generalized free energy as black hole solutions, it seems that black hole solutions at $z = \hat{z}$ (corresponding to $S = \hat{S}$) can exist for any temperature. However, from the perspective of the spacetime geometry near the event horizon, only $\tau = \hat{\tau}$ can correspond to some Euclidean spacetime which can avoid conical singularities, while $\tau > \hat{\tau}$ or $\tau < \hat{\tau}$ will corresponds to the spacetime with conical singularities. Therefore, we believe that $z = \hat{z}$ state can exist only for $\tau = \hat{\tau}$, and its stability depends on the branch. The state $z=\hat{z}$ of branch 2 is stable, while the phase state $z=\hat{z}$ of the branch 1 is unstable.

\section{The case of $\nu =10$ and $\tau = 1$} \label{A_Phiphi100}

\begin{figure}
	\centering
	\includegraphics[width=5 cm]{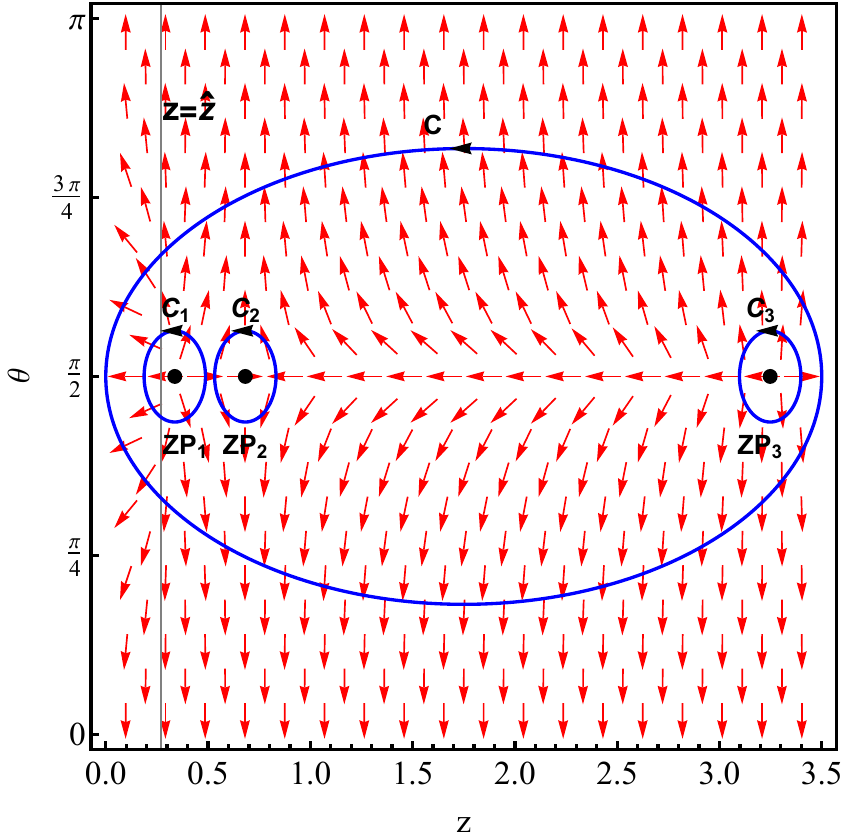}
	\caption{ The vector field constructed by $\phi$ with $\ell_3 = 1/\pi$, $\mathcal{G}_3 = 1$, $\nu = 10$, and $\tau = 1$. The zero points marked with black dots are approximately at $(z,\theta) = (0.34,\pi/2), (0.68,\pi/2)$, and $(3.25,\pi/2)$ for $ZP_1$, $ZP_2$ and $ZP_3$, respectively. This light gray thin line represents the boundary $z=\hat{z} \approx 0.27$. The blue contours $C_i$ are closed loops enclosing the zero points $ZP_i$, and the contour $C$ enclose all the zero points. Moreover, the heat capacity for each zero points are $C_{ZP_1} \approx -0.137, C_{ZP_2} \approx  0.356$, and $C_{ZP_3} \approx  -0.015$.}
	\label{FIG_phiz100}
\end{figure}

\begin{figure}
	\centering
	\includegraphics[width=5 cm]{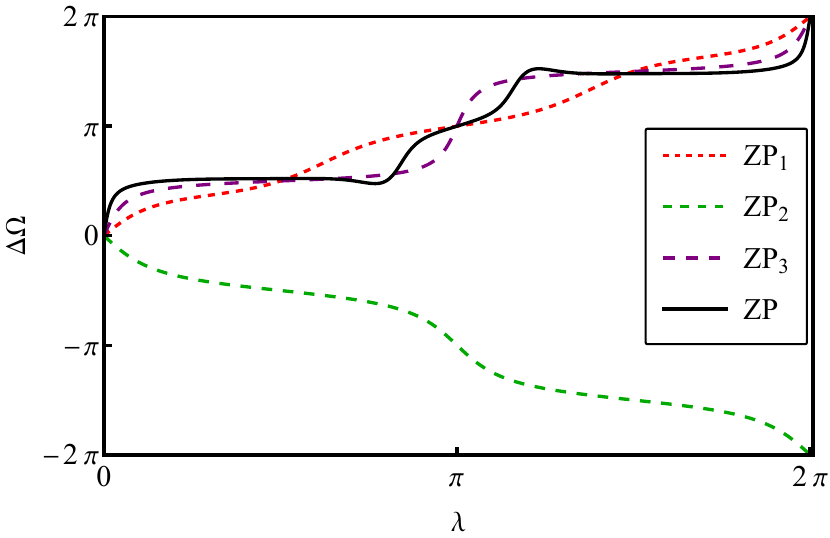}
	\caption{ The curve integral of vector along the different contours. And the winding number can be given by Eq.~\eqref{eq_windingnumber}. These contours integrals correspond to the curve in Fig.~\ref{FIG_phiz100}. Here, $\lambda$ is the angle parameter of the curve with respective to corresponding black point in Fig.~\ref{FIG_phiz100}, and $\Delta\Omega$ is the angle change of the vector field $\phi$ along this curve. Obviously, the winding number of each zero points are given by $w_{ZP_1} = 1, w_{ZP_2} = -1$, and $w_{ZP_3} = 1$. } \label{FIG_phiz100Curve}
\end{figure}

In this Appendix, we present a figure illustrating the vector field pattern \eqref{eq_mapphi} in Fig.~\ref{FIG_phiz100} with the parameters $\ell_3 = 1/\pi$, $\mathcal{G}_3 = 1$, $\nu = 10$, and $\tau = 1$. Furthermore, the vector field is shown in Fig.~\ref{FIG_phiz100Curve}. By examining these figures, we can readily identify three zero points associated with the saddle points, winding numbers, and stability, as follows: $ZP_1$ corresponds to an extreme point with $w = 1$ and is unstable; $ZP_2$ represents a saddle point with $w = -1$ and is stable; $ZP_3$ corresponds to another extreme point with $w = 1$ and is unstable. The relationship between the winding number and stability is listed in TABLE.~\ref{Table_ThermStabilityWindingN}. However, these findings deviate from those presented in Ref.~\cite{Wei:2022BHasTopDefect}, indicating an anomalous behavior.


\begin{figure}
	\begin{center}
		\subfigure[ Foliation 2: The vector field $\Phi_2$ with $0<S<\hat{S}$.	\label{FIG_Phi2S100}]{\includegraphics[width=4cm]{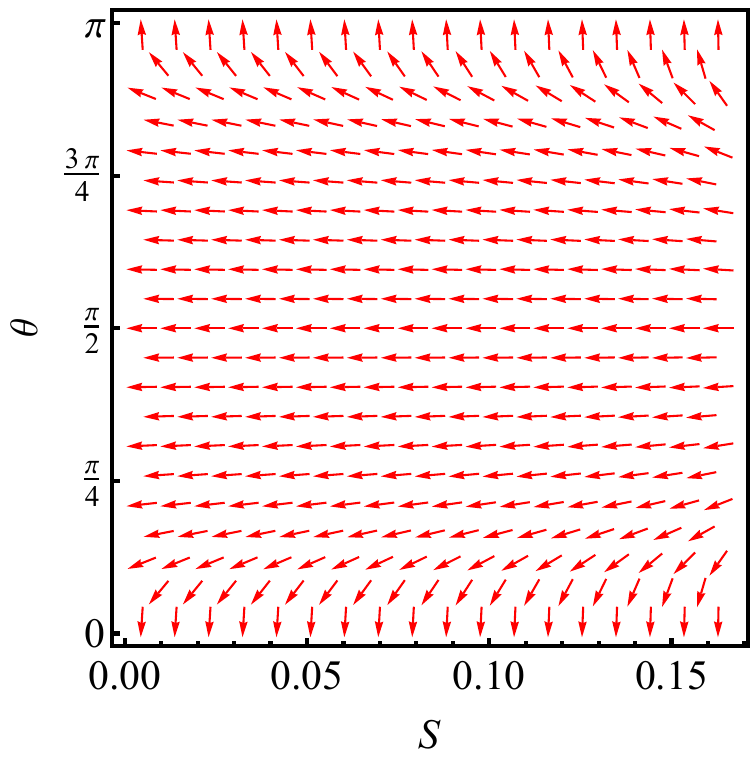}}
		\ \ \ \ \
		\subfigure[ Foliation 1: The vector field $\Phi_1$ with $0<S<\hat{S}$. \label{FIG_Phi1S100}]{\includegraphics[width=4cm]{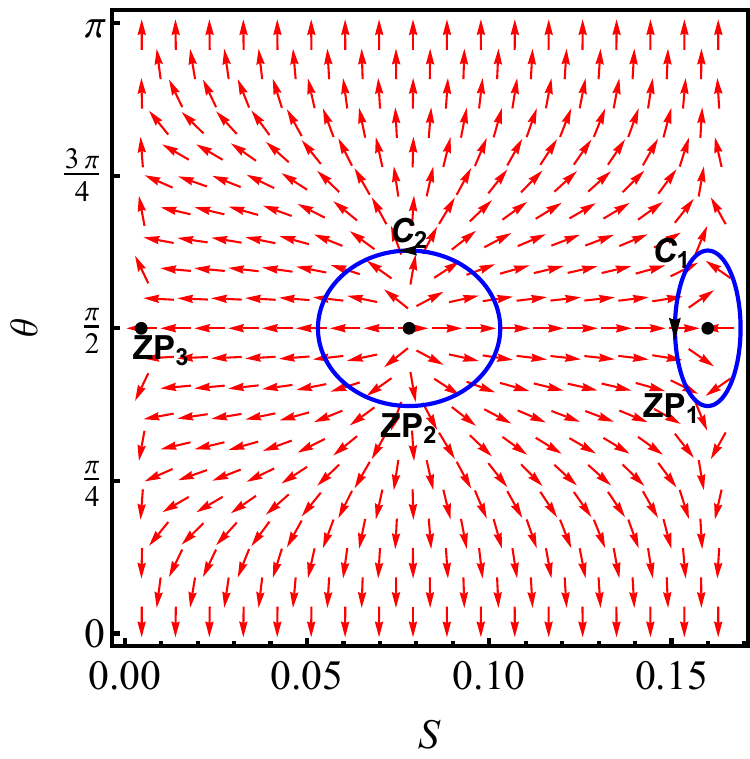}}
		\ \ \ \ \
		\subfigure[ Foliation 1: The vector field $\Phi_1$ with $0<S<0.01$. 	\label{FIG_Phi1S100L}]{\includegraphics[width = 4cm]{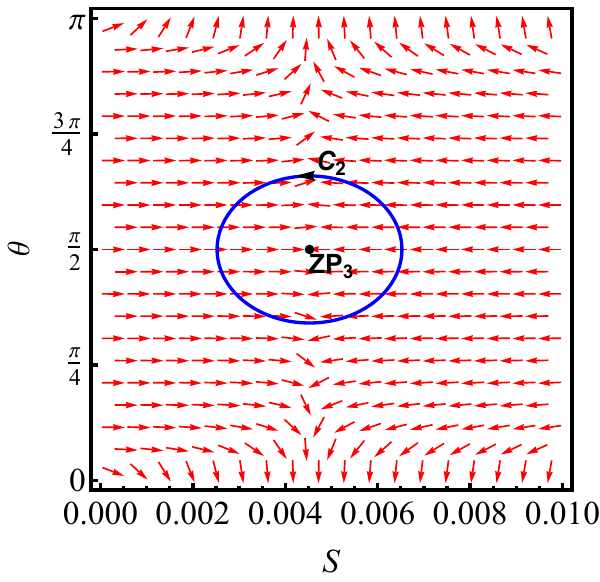}}
	\end{center}
	\caption{The vector field constucted by $\Phi$ with $\ell_3 = 1/\pi$, $\mathcal{G}_3 = 1$, $\nu = 10$, and $\tau = 1$. (c) is an enlarged view of (b). The zero points marked with black dots are approximately at $(S,\theta) = (-0.160,\pi/2)$, $(0.078,\pi/2)$, and $(-0.005,\pi/2)$ for $ZP_1$, $ZP_2$ and $ZP_3$, respectively. These points correspond to the points in Fig.~\ref{FIG_phiz100}.}
	\label{FIG_Phi100}
\end{figure}

The vector field pattern described by Eqs.~\eqref{eq_Phi1def} and \eqref{eq_Phi2def} is plotted in Fig.~\ref{FIG_Phi100}. This plot reveals three zero points, namely $ZP_1$, $ZP_2$, and $ZP_3$, which correspond to the three points shown in Fig.~\ref{FIG_phiz100}. Notably, these zero points correspond to the characteristics of saddle points or extreme points, winding numbers, and stability. Specifically, $ZP_1$ corresponds to a saddle point with $w = -1$ and is unstable. $ZP_2$ represents an extreme point with $w = 1$ and is stable. Lastly, $ZP_3$ corresponds to another saddle point with $w = -1$ and is unstable. Therefore, the winding numbers provided by $\Phi$ offer a meaningful physical explanation.

\end{document}